\documentclass[12pt,a4paper]{article}

\usepackage[margin=1in]{geometry}
\usepackage{amsmath,amssymb,amsthm,mathtools,bm}
\usepackage{graphicx}
\usepackage{booktabs}
\usepackage{multirow}
\usepackage{adjustbox}
\usepackage{xcolor}
\usepackage{url}
\usepackage{float}
\usepackage{subcaption}
\usepackage[round]{natbib}
\usepackage[hidelinks]{hyperref}
\usepackage{algorithm}
\usepackage{algorithmic}
\usepackage{tikz}
\usetikzlibrary{shapes}

\numberwithin{equation}{section}

\theoremstyle{plain}
\newtheorem{theorem}{Theorem}

\newtheorem{corollary}{Corollary}
\newtheorem{definition}{Definition}

\newcommand{\spanop}{\operatorname{span}}

\newcommand{\bM}{\bm{M}}
\newcommand{\bV}{\bm{V}}

\newcommand{\bI}{\bm{I}}
\newcommand{\bPhi}{\bm{\Phi}}
\newcommand{\bDelta}{\bm{\Delta}}
\newcommand{\bSigma}{\bm{\Sigma}}

\newcommand{\bbeta}{\bm{\beta}}

\newcommand{\boldeta}{\bm{\eta}}

\newcommand{\indep}{\perp\!\!\!\perp}

\title{Groupwise Predictor Envelope Models for Multivariate Linear Regression}

\author{
  Sota Osumi\textsuperscript{*},
  Akira Okazaki\textsuperscript{$\dagger$},
  Shuichi Kawano\textsuperscript{$\ddagger$}\\[1ex]
  \small \textsuperscript{*}Graduate School of Mathematics, Kyushu University\\
  \small \textsuperscript{$\dagger$}The Institute of Statistical Mathematics\\
  \small \textsuperscript{$\ddagger$}Faculty of Mathematics, Kyushu University
}

\date{}

\begin{document}

\maketitle

\begin{abstract}
Envelope methods improve estimation efficiency in multivariate analysis by isolating low-dimensional structures that contain all the information material to the parameter of interest. In multivariate linear regression with random predictors, predictor envelope models achieve this goal by removing variation in the predictors that is immaterial to the regression. In many applications, observations are naturally divided into several groups, such as treatment groups, regions, or demographic strata, and the regression relationship may differ between groups. Motivated by this setting, we propose a groupwise predictor envelope model for multivariate linear regression. The proposed model assumes that the group-specific regression coefficient matrices are represented through a common predictor envelope subspace while allowing group-specific regression effects and group-specific error covariance matrices. We derive an objective function for estimating the common predictor envelope, obtain the corresponding regression estimators, and establish asymptotic normality together with an explicit asymptotic variance formula. Moreover, we show that the proposed estimator is asymptotically more efficient than the estimator obtained by fitting predictor envelope models separately to each group. This theoretical advantage over the existing work is also demonstrated through simulation studies.
\end{abstract}


\section{Introduction}

Envelope methods are designed to improve estimation efficiency by identifying a low-dimensional structure that contains all variation material to the target parameter while excluding immaterial variation. Since the original development of envelope methodology, envelope-based ideas have been extended to a variety of multivariate problems \citep{cook2007dimension,CookLiChiaromonte2010,Cook2018,Lee2020,intro2018}. In multivariate linear regression, two representative examples are the response envelope model and the predictor envelope model. The former improves efficiency by removing response variation that is immaterial to the regression coefficients, while the latter exploits the covariance structure of random predictors to remove predictor variation that is irrelevant to the regression relationship \citep{CookLiChiaromonte2010,Cook2013PLS,Cook2018}.

In many practical situations, the data are naturally partitioned into several groups. We refer to such data as multi-group data.  Typical examples include sex-specific analyses, regional studies, treatment groups, and data collected under different operating conditions. In such settings, the regression relationship may differ across groups, and it is natural to allow group-specific regression coefficients. A straightforward strategy is to fit a separate regression model within each group. However, this approach ignores structures shared across groups and may therefore lead to a loss of efficiency. To address this issue, \citet{park2017groupwise} proposed a response envelope model for multi-group data. Their model assumes a common envelope subspace across groups, thus integrating group-specific information in the estimation procedure. They showed that this approach yields efficiency gains relative to fitting separate envelope models for each group.

This paper extends the predictor envelope model to multi-group data. Specifically, we consider multivariate linear regression with random predictors and propose a \emph{groupwise predictor envelope model} that assumes a common predictor envelope subspace across groups. This formulation extends the predictor envelope framework of \citet{Cook2013PLS} to multi-group data while following the structural logic of the groupwise envelope model in \citet{park2017groupwise}. The key idea is to estimate the material predictor subspace by integrating information across groups, while still allowing group-specific regression effects and group-specific response error covariance matrices.

The main contributions of this paper are as follows. First, we formulate the groupwise predictor envelope model and characterize its common predictor envelope through a collection of group-specific predictor covariance matrices. Second, we derive a likelihood-based objective function for estimating the common envelope subspace on a Grassmann manifold. Third, we establish asymptotic normality and an explicit asymptotic variance formula, and further we show that the proposed estimator is asymptotically more efficient than the estimator obtained by fitting predictor envelope models separately to each group.

The remainder of the paper is organized as follows. Section~\ref{sec:preliminaries} reviews reducing subspaces and predictor envelopes. Section~\ref{sec:method} introduces the groupwise predictor envelope model, describes its estimation procedure, and establishes its asymptotic properties. Section~\ref{sec:simulation} presents simulation studies and a real data analysis to evaluate the performance of the proposed method. Section~\ref{sec:conclusion} concludes with a brief discussion.

We use the following notation throughout this paper. Let $\mathbb{S}^{p \times p}$ denote the set of all $p \times p$ symmetric matrices. For a matrix $\mathbf{A}$, $\mathbf{A}^{\dagger}$ denotes the Moore--Penrose inverse of $\mathbf{A}$, $\operatorname{vec}(\mathbf{A})$ denotes the vector obtained by stacking the columns of $\mathbf{A}$, and $\operatorname{span}(\mathbf{A})$ denotes the subspace spanned by the columns of $\mathbf{A}$. The Kronecker product is denoted by $\otimes$. A basis matrix for a subspace $\mathcal{S}$ is any matrix whose columns form a basis for $\mathcal{S}$. For a subspace $\mathcal{S} \subseteq \mathbb{R}^p$, let $\mathbf{P}_{\mathcal{S}}$ denote the orthogonal projection onto $\mathcal{S}$ and let $\mathbf{Q}_{\mathcal{S}}=\mathbf{I}_p-\mathbf{P}_{\mathcal{S}}$ denote the projection onto its orthogonal complement. The notation $\indep$ means that two random variables are independent, and $\xrightarrow{d}$ denotes convergence in distribution.

\section{Preliminaries}\label{sec:preliminaries}
In this section, we introduce the basic concepts of envelopes. Section~\ref{sec:review-envelope} gives the definition of envelopes. Section~\ref{sec:predictor-envelope} presents the formulation of the predictor envelope model.

\subsection{Review of Envelopes}\label{sec:review-envelope}

We first review the notions of reducing subspaces and envelopes, which form the basis of envelope methodology.

\begin{definition}
Let $\mathbf{M} \in \mathbb{S}^{p \times p}$ and let $\mathcal{R} \subseteq \mathbb{R}^p$ be a subspace. Then $\mathcal{R}$ is called a reducing subspace of $\mathbf{M}$ if
\begin{align}
\mathbf{M}
=
\mathbf{P}_{\mathcal{R}} \mathbf{M} \mathbf{P}_{\mathcal{R}}
+
\mathbf{Q}_{\mathcal{R}} \mathbf{M} \mathbf{Q}_{\mathcal{R}}.
\end{align}
\end{definition}

This notion is standard in linear algebra (see \citet{conway1990course}, p. 38). It should be distinguished from the usual notion of dimension reduction in statistics. In envelope methodology, reducing subspaces provide the mathematical foundation for defining an envelope.

The following definition of an envelope was introduced by \citet{CookLiChiaromonte2010}. An envelope is a low-dimensional subspace that retains all the information in the data relevant to the parameter of interest.

\begin{definition}
Let $\mathbf{M} \in \mathbb{S}^{p \times p}$ and let $\mathcal{B} \subseteq \operatorname{span}(\mathbf{M})$ be a subspace. Then the \emph{$\mathbf{M}$-envelope} of $\mathcal{B}$, denoted by $\mathcal{E}_{\mathbf{M}}(\mathcal{B})$, is defined as the intersection of all reducing subspaces of $\mathbf{M}$ that contain $\mathcal{B}$.
\end{definition}

Within the regression framework, $\mathcal{E}_{\mathbf{M}}(\mathcal{B})$ serves as a low-dimensional subspace that contains all information relevant to the parameter of interest while excluding immaterial variation.

\subsection{Predictor Envelope Model}\label{sec:predictor-envelope}
To describe the relationship between a response vector
$\bm{Y} \in \mathbb{R}^r$ and a predictor vector
$\bm{X} \in \mathbb{R}^p$, we consider the multivariate linear regression model
\begin{align}
\bm{Y} = \bm{\mu} + \bm{\beta} \left(\bm{X}-\bm{\mu}_{\bm{X}} \right) + \bm{\varepsilon}.
\label{eq:multivariate_regression}
\end{align}
Here, $\bm{\mu} \in \mathbb{R}^r$ is an intercept vector, $\bm{\beta} \in \mathbb{R}^{r \times p}$ is a regression coefficient matrix, and $\bm{\varepsilon} \in \mathbb{R}^r$ is an error vector. We assume that $\bm{\varepsilon} \sim N_r(\bm{0}, \bm{\Sigma})$ and
$\bm{X} \sim N_p(\bm{\mu}_{\bm{X}}, \bm{\Sigma}_{\bm{X}})$, where
$\bm{\Sigma} \in \mathbb{R}^{r \times r}$ is the covariance matrix of
$\bm{\varepsilon}$, and $\bm{\mu}_{\bm{X}} \in \mathbb{R}^p$ and
$\bm{\Sigma}_{\bm{X}} \in \mathbb{R}^{p \times p}$ are the mean vector and
covariance matrix of $\bm{X}$, respectively. Furthermore, we assume that
$\bm{X}$ and $\bm{\varepsilon}$ are independent.

To introduce the predictor envelope model, \citet{Cook2013PLS} first impose the following two conditions on the multivariate linear regression model \eqref{eq:multivariate_regression}:
\begin{align}
    (1) \: \mathrm{cov}(\bm{Y}, \bm{\Phi}_0^{\top} \bm{X} \mid \bm{\Phi}^{\top} \bm{X})=0,
\qquad
(2) \: \mathrm{cov}(\bm{\Phi}^{\top} \bm{X}, \bm{\Phi}_0^{\top} \bm{X})=0.
\label{eq:TwoConditions}
\end{align}
Here, $(\bm{\Phi},\bm{\Phi}_0)\in\mathbb{R}^{p\times p}$ is an orthogonal matrix and
$\bm{\Phi}\in\mathbb{R}^{p\times q}$ with $q\le p$. Let
$\mathcal{W}_q := \mathrm{span}(\bm{\Phi})$. As shown in \citet{Cook2013PLS}, it is known that the two
conditions \eqref{eq:TwoConditions} are equivalent to the following conditions:
\begin{align}\label{condtion:env}
\text{(a)}\:\mathrm{span}(\bm{\beta}^{\top})\subseteq \mathcal{W}_q,
\qquad
\text{(b)}\:
\bm{\Sigma_X}= \mathbf{P}_{\mathcal{W}_q}\bm{\Sigma_X}\mathbf{P}_{\mathcal{W}_q}+
\mathbf{Q}_{\mathcal{W}_q}\bm{\Sigma_X} \mathbf{Q}_{\mathcal{W}_q}.
\end{align}
The conditions (a) and (b) hold if and only if $\mathcal{W}_q$ is a reducing subspace of
$\bm{\Sigma_X}$ that contains $\mathrm{span}(\bm{\beta}^{\top})$. The $\bm{\Sigma_X}$-envelope of $\mathrm{span}(\bm{\beta}^{\top})$ is defined as the smallest reducing subspace that satisfies these
properties and is denoted by $\mathcal{E}_{\bm{\Sigma_X}}\{\mathrm{span}(\bm{\beta}^{\top})\}$. 

Under condition~(a), there exists a matrix $\bm{\eta} \in \mathbb{R}^{q \times r}$ such that $\bm{\beta}^{\top} = \bm{\Phi}\bm{\eta}$. Moreover, since $\mathbf{P}_{\mathcal{W}_q} = \bm{\Phi}\bm{\Phi}^{\top}$ and $\mathbf{Q}_{\mathcal{W}_q} = \bm{\Phi}_0\bm{\Phi}_0^{\top}$, condition~(b) implies that
\[
\bm{\Sigma}_{\bm{X}} = \bm{\Phi}\left(\bm{\Phi}^{\top}\bm{\Sigma}_{\bm{X}}\bm{\Phi}\right)\bm{\Phi}^{\top} + \bm{\Phi}_0\left(\bm{\Phi}_0^{\top}\bm{\Sigma}_{\bm{X}}\bm{\Phi}_0\right)\bm{\Phi}_0^{\top}.
\]
Therefore, under the above assumptions, \citet{Cook2013PLS} formulate the predictor envelope model as follows:
\begin{align}
\bm{Y} = \bm{\mu} + \bm{\eta}^{\top}\bm{\Phi}^{\top}(\bm{X}-\bm{\mu}_{\bm{X}}) + \bm{\varepsilon}, \quad 
\bm{\Sigma}_{\bm{X}} = \bm{\Phi}\bm{\Delta}\bm{\Phi}^{\top} + \bm{\Phi}_0\bm{\Delta}_0\bm{\Phi}_0^{\top},
\end{align}
where $\bm{\Delta} = \bm{\Phi}^{\top}\bm{\Sigma}_{\bm{X}}\bm{\Phi}$ and $\bm{\Delta}_0 = \bm{\Phi}_0^{\top}\bm{\Sigma}_{\bm{X}}\bm{\Phi}_0$ are positive definite matrices.

We briefly summarize the likelihood-based derivation of the predictor envelope objective function given by \citet{Cook2013PLS, intro2018}.
To describe the derivation of the objective function, let $\mathbf{C}=(\bm{X}^{\top},\bm{Y}^{\top})^{\top}$,
and let $\bm{\Sigma}_{\mathbf{C}}$ and $\bm{S}_{\mathbf{C}}$ denote the covariance matrix and the sample covariance matrix of $\mathbf{C}$, respectively. 
For a given value of $q$, estimation is carried out by minimizing the objective function $F_q(\bm{S}_{\mathbf{C}}, \bm{\Sigma}_{\mathbf{C}}) = \log |\bm{\Sigma}_{\mathbf{C}}| + \mathrm{tr} (\bm{S}_{\mathbf{C}} \bm{\Sigma}_{\mathbf{C}}^{-1})$, which is derived from the multivariate normal log-likelihood with the population mean vector replaced by the sample mean vector.
Under the multivariate linear regression model \eqref{eq:multivariate_regression}, $\bm{\Sigma}_{\mathbf{C}}$ is given by
\begin{align}
\bm{\Sigma}_{\mathbf{C}}= 
\begin{pmatrix}
    \bm{\Sigma}_{\bm{X}} & \bm{\Sigma}_{\bm{X,Y}} \\
    \bm{\Sigma}_{\bm{X,Y}}^{\top} & \bm{\Sigma}_{\bm{Y}}
\end{pmatrix}
=
\begin{pmatrix}\bm{\Sigma}_{\bm{X}} & \bm{\Sigma}_{\bm{X}}\bm{\beta}^{\top}\\ \bm{\beta}\bm{\Sigma}_{\bm{X}} & \bm{\beta}\bm{\Sigma}_{\bm{X}}\bm{\beta}^{\top}+\bm{\Sigma}\end{pmatrix},
\end{align}
where $\bm{\Sigma}_{\bm{X}}$ and $\bm{\Sigma}_{\bm{Y}}$ denote the covariance matrices of $\bm{X}$ and $\bm{Y}$, respectively, and $\bm{\Sigma}_{\bm{X},\bm{Y}}$ denotes the covariance matrix of $\bm{X}$ and $\bm{Y}$.
Using the parameterizations $\bm{\beta}^{\top}=\bm{\Phi}\bm{\eta}$ and $\bm{\Sigma}_{\bm{X}}=\bm{\Phi}\bm{\Delta}\bm{\Phi}^{\top}+\bm{\Phi}_0\bm{\Delta}_0\bm{\Phi}_0^{\top}$, we obtain
\begin{align}\label{eq:variance_of_C}
\bm{\Sigma}_{\mathbf{C}}&=\begin{pmatrix}\bm{\Phi}\bm{\Delta}\bm{\Phi}^{\top}+\bm{\Phi}_0\bm{\Delta}_0\bm{\Phi}_0^{\top} & \bm{\Phi}\bm{\Delta}\bm{\eta}\\ \bm{\eta}^{\top}\bm{\Delta}\bm{\Phi}^{\top} & \bm{\eta}^{\top}\bm{\Delta}\bm{\eta}+\bm{\Sigma}\end{pmatrix} \notag\\
&= \begin{pmatrix}
    \bm{\Phi} & \bm{\Phi}_0 & 0\\
    0 & 0 & \bm{I}_r
\end{pmatrix}
\begin{pmatrix}
    \bm{\Delta} & 0 &\bm{\Delta\eta} \\
    0 & \bm{\Delta}_0 & 0\\
    \bm{\eta}^{\top}\bm{\Delta} & 0 & \bm{\Sigma}_{\bm{Y}} 
\end{pmatrix}
\begin{pmatrix}
    \bm{\Phi}^{\top} & 0 \\
    \bm{\Phi}_0^{\top} & 0 \\
    0 & \bm{I}_r
\end{pmatrix} \notag\\
&= {\mathbf{O}}\bm{\Sigma}_{\mathbf{O}^{\top}\mathbf{C}}\mathbf{O}^{\top},
\end{align}
where 
\[
\mathbf{O} = \begin{pmatrix}
    \bm{\Phi} & \bm{\Phi}_0 & 0\\
    0 & 0 & \bm{I}_r
\end{pmatrix} \in \mathbb{R}^{(p+r) \times (p+r)}
\]
is an orthogonal matrix and 
\[
\bm{\Sigma}_{\mathbf{O}^{\top}\mathbf{C}} = \begin{pmatrix}
    \bm{\Delta} & 0 & \bm{\Delta}\bm{\eta} \\
    0 & \bm{\Delta}_0 & 0 \\
    \bm{\eta}^{\top} \bm{\Delta} & 0 & \bm{\Sigma}_{\bm{Y}}
\end{pmatrix} \in \mathbb{R}^{(p+r) \times (p+r)}
\]
is the covariance matrix of the transformed vector $\mathbf{O}^{\top}\mathbf{C}$. 
The objective function can therefore be expressed in terms of the five parameters $\bm{\Phi}, \bm{\Delta}, \bm{\Delta}_0, \bm{\eta}$, and $\bm{\Sigma}_{\bm{Y}}$, which jointly determine $\bm{\Sigma}_{\mathbf{C}}$. 

Define the permutation matrix
\[
P
=
\begin{pmatrix}
\mathbf{I}_q & \mathbf{0} & \mathbf{0}\\
\mathbf{0} & \mathbf{0} & \mathbf{I}_{p-q}\\
\mathbf{0} & \mathbf{I}_r & \mathbf{0}
\end{pmatrix}.
\]
Then,
\[
P^{\top}
\bm{\Sigma}_{\mathbf{O}^{\top}\mathbf{C}}
P
=
\begin{pmatrix}
\bm{\Delta}
&
\bm{\Delta}\bm{\eta}
&
\mathbf{0}
\\
\bm{\eta}^{\top}\bm{\Delta}
&
\bm{\Sigma}_{\mathbf{Y}}
&
\mathbf{0}
\\
\mathbf{0}
&
\mathbf{0}
&
\bm{\Delta}_0
\end{pmatrix}
=
\begin{pmatrix}
\bm{A} & \mathbf{0}\\
\mathbf{0} & \bm{\Delta}_0
\end{pmatrix},
\qquad
\bm{A}
=
\begin{pmatrix}
\bm{\Delta}
&
\bm{\Delta}\bm{\eta}
\\
\bm{\eta}^{\top}\bm{\Delta}
&
\bm{\Sigma}_{\mathbf{Y}}
\end{pmatrix}.
\]
Since \(P\) is orthogonal,
\[
\begin{aligned}
\log
\left|
\bm{\Sigma}_{\mathbf{O}^{\top}\mathbf{C}}
\right|
&=
\log
\left|
P^{\top}
\bm{\Sigma}_{\mathbf{O}^{\top}\mathbf{C}}
P
\right|
\\
&=
\log|\bm{A}|
+
\log|\bm{\Delta}_0|.
\end{aligned}
\]
Moreover,
\[
\left(
P^{\top}
\bm{\Sigma}_{\mathbf{O}^{\top}\mathbf{C}}
P
\right)^{-1}
=
\begin{pmatrix}
\bm{A}^{-1} & \mathbf{0}\\
\mathbf{0} & \bm{\Delta}_0^{-1}
\end{pmatrix}.
\]
Let \(\mathbf{Z}=\bigl((\bm{\Phi}^{\top}\mathbf{X})^{\top},\mathbf{Y}^{\top}\bigr)^{\top}\), and let \(\bm{S}_{\mathbf{Z}}\) denote its sample covariance matrix. Using the invariance of the trace under cyclic permutations, we obtain
\[
\begin{aligned}
\operatorname{tr}
\left(
\bm{S}_{\mathbf{O}^{\top}\mathbf{C}}
\bm{\Sigma}_{\mathbf{O}^{\top}\mathbf{C}}^{-1}
\right)
&=
\operatorname{tr}
\left[
\left(
P^{\top}
\bm{S}_{\mathbf{O}^{\top}\mathbf{C}}
P
\right)
\left(
P^{\top}
\bm{\Sigma}_{\mathbf{O}^{\top}\mathbf{C}}
P
\right)^{-1}
\right]
\\
&=
\operatorname{tr}
\left(
\bm{S}_{\mathbf{Z}}\bm{A}^{-1}
\right)
+
\operatorname{tr}
\left(
\bm{\Phi}_0^{\top}
\bm{S}_{\mathbf{X}}
\bm{\Phi}_0
\bm{\Delta}_0^{-1}
\right).
\end{aligned}
\]
Consequently,
\[
\begin{aligned}
F_q(\bm{S}_{\mathbf{C}},\bm{\Sigma}_{\mathbf{C}})
={}&
\left\{
\log|\bm{A}|
+
\operatorname{tr}
\left(
\bm{S}_{\mathbf{Z}}\bm{A}^{-1}
\right)
\right\}
\\
&+
\left\{
\log|\bm{\Delta}_0|
+
\operatorname{tr}
\left(
\bm{\Phi}_0^{\top}
\bm{S}_{\mathbf{X}}
\bm{\Phi}_0
\bm{\Delta}_0^{-1}
\right)
\right\}.
\end{aligned}
\]
Therefore, for fixed \(\bm{\Phi}\), the objective function separates into one term involving only \(\bm{\Delta}_0\) and another involving only \(\bm{\Delta}\), \(\bm{\eta}\), and \(\bm{\Sigma}_{\mathbf{Y}}\), so that these two sets of parameters can be minimized separately.
The first term is minimized at \(\bm{A}=\bm{S}_{\mathbf{Z}}\), whereas the second term is minimized at \(\bm{\Delta}_0=\bm{\Phi}_0^{\top}\bm{S}_{\bm{X}}\bm{\Phi}_0\). Equating the corresponding blocks of \(\bm{A}\) and \(\bm{S}_{\mathbf{Z}}\), we obtain
\[
\bm{\Sigma}_{\bm{Y}}
=
\bm{S}_{\bm{Y}},
\quad
\bm{\Delta}
=
\bm{\Phi}^{\top}\bm{S}_{\bm{X}}\bm{\Phi},
\quad
\bm{\Delta}_0
=
\bm{\Phi}_0^{\top}\bm{S}_{\bm{X}}\bm{\Phi}_0,
\quad
\bm{\eta}
=
\left(
\bm{\Phi}^{\top}\bm{S}_{\bm{X}}\bm{\Phi}
\right)^{-1}
\bm{\Phi}^{\top}\bm{S}_{\bm{X},\bm{Y}}.
\]
Here, \(\bm{S}_{\bm{X}}\) and \(\bm{S}_{\bm{Y}}\) denote the sample covariance matrices of \(\bm{X}\) and \(\bm{Y}\), respectively, and \(\bm{S}_{\bm{X},\bm{Y}}\) denotes their sample cross-covariance matrix.
Substituting these expressions into $F_q(\bm{S}_{\mathrm{\bm{C}}},\bm{\Sigma}_{\mathrm{C}})$ leads to the following estimator of the envelope for a given value of $q$:
\begin{equation}
\widehat{\mathcal{E}}_{\bm{\Sigma}_{\bm X}}
\left\{
\operatorname{span}(\bm{\beta}^{\top})
\right\}
=
\underset{\operatorname{span}(\bm{\Phi})\in\mathcal{G}(q,p)}
{\operatorname{argmin}}
\left\{
\log\left|
\bm{\Phi}^{\top}
\bm{S}_{\bm{X}\mid\bm{Y}}
\bm{\Phi}
\right|
+
\log\left|
\bm{\Phi}^{\top}
\bm{S}_{\bm{X}}^{-1}
\bm{\Phi}
\right|
\right\},
\end{equation}
where $\bm{S}_{\bm{X}\mid\bm{Y}}$ denotes the residual covariance matrix from the regression of $\bm{X}$ on $\bm{Y}$.
Furthermore, the estimator
of the regression coefficient matrix in the predictor envelope model is given by
\begin{align}
\widehat{\bm{\beta}}_{\mathrm{env}}
=
\widehat{\bm{\beta}}_{\mathrm{OLS}}
\mathbf{P}^{\top}_{\widehat{\mathcal{E}}(S_{\bm{X}})} ,
\end{align}
where $\widehat{\bm{\beta}}_{\mathrm{OLS}}$ is the ordinary least squares estimator,
and $\mathbf{P}_{\widehat{\mathcal{E}}(S_{\bm{X}})}$ is the projection matrix onto the $\widehat{\mathcal{E}}_{\bm{\Sigma_X}}(\mathrm{span}(\bm{\beta}^{\top}))$ with respect to the $S_{\bm X}$-inner product. 

Using Proposition 4.1 of \citet{Shapiro1986}, \citet{Cook2013PLS} showed that
$\sqrt{n}(\widehat{\bm{\beta}}_{\mathrm{env}}-\bm{\beta})$ is asymptotically
normal with mean zero vector and covariance matrix $V_{\mathrm{env}}$ given by
\begin{align}
V_{\mathrm{env}}
&=
\bm{\Phi} \bm{\Delta}^{-1}\bm{\Phi}^{\top} \otimes \bm{\Sigma}
+
(\bm{\Phi}_0 \otimes \bm{\eta}^{\top}) T^\dagger (\bm{\Phi}_0^{\top} \otimes \bm{\eta}),
\end{align}
where $T =
\bm{\Delta}_0 \otimes \bm{\eta} \bm{\Sigma}^{-1}\bm{\eta}^{\top}
+
\bm{\Delta}_0^{-1}\otimes \bm{\Delta}
+
\bm{\Delta}_0 \otimes \bm{\Delta}^{-1}
-
2 \bm{I}_{p-q}\otimes \bm{I}_q .
$
On the other hand, the asymptotic variance of the ordinary least squares estimator is given by
\begin{align}
V_{\mathrm{OLS}}
=
\bm{\Sigma}_{\bm{X}}^{-1}\otimes\bm{\Sigma}
=
\bm{\Phi}\bm{\Delta}^{-1}\bm{\Phi}^{\top}\otimes\bm{\Sigma}
+
\bm{\Phi}_0\bm{\Delta}_0^{-1}\bm{\Phi}_0^{\top}\otimes\bm{\Sigma},
\end{align}
and \citet{Cook2013PLS} showed that:
\begin{align}
V_{\mathrm{OLS}}\succeq V_{\mathrm{env}}.
\end{align}
This implies that the predictor envelope estimator never performs worse asymptotically than the ordinary least squares estimator under the condition \eqref{eq:TwoConditions}.

\section{Proposed Method}\label{sec:method}

In this section, we propose a groupwise predictor envelope model for multivariate
linear regression. The proposed model extends the predictor envelope model
in Section~\ref{sec:predictor-envelope} to groupwise data by introducing
a common envelope subspace shared across groups. Section~\ref{sec:model-formulation} presents the model formulation. Section~\ref{sec:estimation} describes the
estimation procedure. Section~\ref{sec:theory} establishes the theoretical properties of the proposed method. 

\subsection{Model formulation}\label{sec:model-formulation}

Let $L$ denote the number of groups, let $n_\ell$ denote the sample size of
group $\ell$, and let $n=\sum_{\ell=1}^L n_\ell$. To describe the
relationship between the response vector
$\bm{Y}_{(\ell)j}\in\mathbb{R}^r$ and the predictor vector
$\bm{X}_{(\ell)j}\in\mathbb{R}^p$ for the $j$th observation in the group $\ell$, we
consider the following multivariate linear regression model extended to
groupwise data:
\begin{align}
\bm{Y}_{(\ell)j}
=
\bm{\mu}_{(\ell)}
+
\bm{\beta}_{(\ell)}
\left(
\bm{X}_{(\ell)j}
-
\bm{\mu}_{\bm{X}_{(\ell)}}
\right)
+
\bm{\varepsilon}_{(\ell)j},
\qquad
j=1,\ldots,n_\ell, \quad \ell=1,\ldots,L.
\end{align}
Here, $\bm{\mu}_{(\ell)}\in\mathbb{R}^r$ is the intercept vector for group $\ell$, $\bm{\mu}_{\bm{X}_{(\ell)}}\in\mathbb{R}^p$ is the mean vector of the predictor variables in group $\ell$, $\bm{\beta}_{(\ell)}\in\mathbb{R}^{r\times p}$ is the regression coefficient matrix for group $\ell$, and $\bm{\varepsilon}_{(\ell)j}\in\mathbb{R}^r$ is the error vector for the $j$th observation in group $\ell$. Furthermore, we assume that
\[
\bm{X}_{(\ell)j} \sim N_p( \bm{\mu}_{\bm{X}_{(\ell)}}, \bm{\Sigma}_{\bm{X}_{(\ell)}} ),
\qquad
\bm{\varepsilon}_{(\ell)j} \sim N_r( \bm{0}, \bm{\Sigma}_{(\ell)} ).
\]
Here, $\bm{\Sigma}_{\bm{X}_{(\ell)}} \in \mathbb{R}^{p \times p}$ is the covariance matrix of $\bm{X}_{(\ell)j}$, and 
$\bm{\Sigma}_{(\ell)} \in \mathbb{R}^{r \times r}$ is the covariance matrix of
$\bm{\varepsilon}_{(\ell)j}$. We also assume that
$\bm{X}_{(\ell)j}$ and $\bm{\varepsilon}_{(\ell)j}$ are independent.

As in the predictor envelope model, we assume that there exists a semi-orthogonal matrix $\bm{\Phi}\in\mathbb{R}^{p\times q}$ satisfying $\bm{\Phi}^{\top}\bm{\Phi}=\bm{I}_q$ such that, for every $\ell=1,\ldots,L$,
\begin{equation}\label{condition:proposed_1}
\textnormal{(1)}\quad \operatorname{cov}\left(\bm{Y}_{(\ell)j},\bm{\Phi}_0^{\top}\bm{X}_{(\ell)j}\mid\bm{\Phi}^{\top}\bm{X}_{(\ell)j}\right)=\bm{0}
\quad
\textnormal{(2)}\quad \operatorname{cov}\left(\bm{\Phi}^{\top}\bm{X}_{(\ell)j},\bm{\Phi}_0^{\top}\bm{X}_{(\ell)j}\right)=\bm{0},
\end{equation}
where $\bm{\Phi}_0\in\mathbb{R}^{p\times(p-q)}$ is an orthonormal complement of $\bm{\Phi}$. 

Let \(\mathcal{E}=\operatorname{span}(\bm{\Phi})\). Since the columns of \(\bm{\Phi}\) and \(\bm{\Phi}_0\) form orthonormal bases of \(\mathcal{E}\) and \(\mathcal{E}^{\perp}\), respectively, the orthogonal projection matrices onto these subspaces are given by $\mathbf{P}_{\mathcal{E}}=\bm{\Phi}\bm{\Phi}^{\top}, 
\mathbf{Q}_{\mathcal{E}}
=
\bm{\Phi}_0\bm{\Phi}_0^{\top}
=
\mathbf{I}_p-\mathbf{P}_{\mathcal{E}}$. 
By the regression model and the independence of $\bm{X}_{(\ell)j}$ and $\bm{\varepsilon}_{(\ell)j}$,
\[
\operatorname{cov}\left(\bm{Y}_{(\ell)j},\bm{\Phi}_0^{\top}\bm{X}_{(\ell)j}\mid\bm{\Phi}^{\top}\bm{X}_{(\ell)j}\right)
=
\bm{\beta}_{(\ell)}\bm{\Phi}_0
\operatorname{var}\left(\bm{\Phi}_0^{\top}\bm{X}_{(\ell)j}\mid\bm{\Phi}^{\top}\bm{X}_{(\ell)j}\right).
\]
Since $\bm{\Sigma}_{\bm{X}_{(\ell)}}$ is positive definite, the conditional covariance matrix on the right-hand side is also positive definite. Hence,
\[
\textnormal{(1)}
\quad\Longleftrightarrow\quad
\bm{\beta}_{(\ell)}\bm{\Phi}_0=\bm{0}
\quad\Longleftrightarrow\quad
\operatorname{span}\left(\bm{\beta}_{(\ell)}^{\top}\right)\subseteq\mathcal{E}.
\]

Moreover,
\[
\operatorname{cov}\left(\bm{\Phi}^{\top}\bm{X}_{(\ell)j},\bm{\Phi}_0^{\top}\bm{X}_{(\ell)j}\right)
=
\bm{\Phi}^{\top}\bm{\Sigma}_{\bm{X}_{(\ell)}}\bm{\Phi}_0.
\]
Therefore,
\[
\textnormal{(2)}
\quad\Longleftrightarrow\quad
\mathbf{P}_{\mathcal{E}}\bm{\Sigma}_{\bm{X}_{(\ell)}}\mathbf{Q}_{\mathcal{E}}
=
\mathbf{Q}_{\mathcal{E}}\bm{\Sigma}_{\bm{X}_{(\ell)}}\mathbf{P}_{\mathcal{E}}
=
\bm{0},
\]
which is equivalent to
\[
\bm{\Sigma}_{\bm{X}_{(\ell)}}
=
\mathbf{P}_{\mathcal{E}}\bm{\Sigma}_{\bm{X}_{(\ell)}}\mathbf{P}_{\mathcal{E}}
+
\mathbf{Q}_{\mathcal{E}}\bm{\Sigma}_{\bm{X}_{(\ell)}}\mathbf{Q}_{\mathcal{E}}.
\]

Thus, conditions~(1) and~(2) are, respectively, equivalent to
\begin{equation}\label{condition:proposed_2}
\textnormal{(a)}\quad \operatorname{span}\left(\bm{\beta}_{(\ell)}^{\top}\right)\subseteq\mathcal{E}
\quad
\textnormal{(b)}\quad
\bm{\Sigma}_{\bm{X}_{(\ell)}}
=
\mathbf{P}_{\mathcal{E}}\bm{\Sigma}_{\bm{X}_{(\ell)}}\mathbf{P}_{\mathcal{E}}
+
\mathbf{Q}_{\mathcal{E}}\bm{\Sigma}_{\bm{X}_{(\ell)}}\mathbf{Q}_{\mathcal{E}},
\end{equation}
for every $\ell=1,\ldots,L$.

We denote this common groupwise predictor envelope by $\mathcal{E}_{\mathcal{M}_{\bm{X}}} (\mathcal{B}')$, where $\mathcal{M}_{\bm{X}}$ denotes the collection of predictor covariance matrices across all groups, and 
\[
\mathcal{B}' = \operatorname{span}(\bm{\beta}^{\top}_{(1)}, \bm{\beta}^{\top}_{(2)}, \ldots,\bm{\beta}^{\top}_{(L)}).
\]
Motivated by this structure, we propose the groupwise predictor envelope model
\begin{equation}\label{sec:proposed formulation}
\bm{Y}_{(\ell)j}
=
\bm{\mu}_{(\ell)}
+
\bm{\eta}_{(\ell)}^{\top}
\bm{\Phi}^{\top}
\left(
\bm{X}_{(\ell)j}
-
\bm{\mu}_{\bm{X}_{(\ell)}}
\right)
+
\bm{\varepsilon}_{(\ell)j}, \ \bm{\Sigma}_{\bm{X}_{(\ell)}} =\bm{\Phi}\bm{\Delta}_{(\ell)}\bm{\Phi}^{\top}+
\bm{\Phi}_0\bm{\Delta}_0\bm{\Phi}_0^{\top}
\end{equation}
for each $\ell = 1, \ldots, L$. 
Here $\bm{\beta}_{(\ell)} = \bm{\eta}^{\top}_{(\ell)} \bm{\Phi}^{\top}$, $\bm{\eta}_{(\ell)} \in \mathbb{R}^{q \times r}$ carries the coordinate of $\bm{\beta}_{(\ell)}$ with respect to $\bm{\Phi}$, and $\bm{\Delta}_{(\ell)} \in \mathbb{R}^{q \times q}$ and $\bm{\Delta}_0 \in \mathbb{R}^{(p-q) \times (p-q)}$ are symmetric matrices that carry the coordinates of $\bm{\Sigma}_{\bm{X}_{(\ell)}}$ with respect to $\bm{\Phi}$ and $\bm{\Phi}_0$, respectively. 
The model allows group-specific covariance structures within the material subspace through $\bm{\Delta}_{(\ell)}$, while imposing a common covariance structure on the immaterial component through $\bm{\Delta}_0$. Thus, the specification accommodates group heterogeneity in the material variation while retaining a shared structure across groups in the immaterial variation.
Note that the groupwise predictor envelope model degenerates to the predictor envelope model in \citet{Cook2013PLS} if $L=1$.  

This model is the predictor-side counterpart of the groupwise envelope model of \citet{park2017groupwise}. Its key feature is that the common subspace $\spanop(\bPhi)$ is estimated using information from all groups, whereas the regression effects and response error covariance matrices remain group-specific.

\subsection{Estimation}\label{sec:estimation}

Let \(n=\sum_{\ell=1}^{L}n_\ell\) and define \(f_\ell=n_\ell/n\) for \(\ell=1,\ldots,L\). For each group, let \(\overline{\bm{X}}_{(\ell)}\) and \(\overline{\bm{Y}}_{(\ell)}\) denote the sample means of the predictors and responses, respectively, and let \(\bm{X}_{(\ell)c}\in\mathbb{R}^{n_\ell\times p}\) and \(\bm{Y}_{(\ell)c}\in\mathbb{R}^{n_\ell\times r}\) denote the corresponding groupwise centered data matrices. Define
\[
\bm{C}_{(\ell)j}
=
\begin{pmatrix}
\bm{X}_{(\ell)j}\\
\bm{Y}_{(\ell)j}
\end{pmatrix},
\]
and let \(\bm{S}_{\bm{C}_{(\ell)}}\) and \(\bm{\Sigma}_{\bm{C}_{(\ell)}}\) denote the sample covariance matrix and population covariance matrix of \(\bm{C}_{(\ell)j}\), respectively. Throughout this subsection, we assume that all sample covariance matrices and projected covariance matrices appearing below are nonsingular.

We estimate the model parameters by minimizing the following weighted discrepancy function:
\begin{equation}\label{eq:groupwise-discrepancy}
F(\bm{S_{C_{(\ell)}}}, \bm{\Sigma_{\bm{C}_{(\ell)}}})
=
\sum_{\ell=1}^{L}
f_\ell
\left[
\log\left|
\bm{\Sigma}_{\bm{C}_{(\ell)}}
\right|
+
\operatorname{tr}
\left\{
\bm{S}_{\bm{C}_{(\ell)}}
\bm{\Sigma}_{\bm{C}_{(\ell)}}^{-1}
\right\}
\right].
\end{equation}
Up to an additive constant and a positive multiplicative factor, \(F\) is equivalent to the negative Gaussian log-likelihood per observation. Normality is used only to motivate the estimation criterion.

For each group, define
\[
\bm{S}_{\bm{X}_{(\ell)}}
=
\frac{1}{n_\ell}
\bm{X}_{(\ell)c}^{\top}
\bm{X}_{(\ell)c},
\qquad
\bm{S}_{\bm{Y}_{(\ell)}}
=
\frac{1}{n_\ell}
\bm{Y}_{(\ell)c}^{\top}
\bm{Y}_{(\ell)c},
\]
and
\[
\bm{S}_{\bm{X}_{(\ell)}\bm{Y}_{(\ell)}}
=
\frac{1}{n_\ell}
\bm{X}_{(\ell)c}^{\top}
\bm{Y}_{(\ell)c},
\qquad
\bm{S}_{\bm{Y}_{(\ell)}\bm{X}_{(\ell)}}
=
\bm{S}_{\bm{X}_{(\ell)}\bm{Y}_{(\ell)}}^{\top}.
\]
The sample covariance matrix of the predictors conditional on the responses is
\[
\bm{S}_{\bm{X}_{(\ell)}\mid\bm{Y}_{(\ell)}}
=
\bm{S}_{\bm{X}_{(\ell)}}
-
\bm{S}_{\bm{X}_{(\ell)}\bm{Y}_{(\ell)}}
\bm{S}_{\bm{Y}_{(\ell)}}^{-1}
\bm{S}_{\bm{Y}_{(\ell)}\bm{X}_{(\ell)}}.
\]
We also define the pooled within-group covariance estimator of the predictors by
\[
\widehat{\bm{\Sigma}}_{\bm{X}}
=
\frac{1}{n}
\sum_{\ell=1}^{L}
\bm{X}_{(\ell)c}^{\top}
\bm{X}_{(\ell)c}
=
\sum_{\ell=1}^{L}
f_\ell
\bm{S}_{\bm{X}_{(\ell)}}.
\]

Under the groupwise predictor envelope model introduced in Section~\ref{sec:model-formulation}, the discrepancy function in \eqref{eq:groupwise-discrepancy} can be minimized explicitly with respect to the nuisance parameters for a fixed \(\bm{\Phi}\). Because the covariance matrix \(\bm{\Delta}_0\) associated with the immaterial subspace is common across groups, its profile estimator is obtained from the pooled within-group covariance estimator \(\widehat{\bm{\Sigma}}_{\bm{X}}\). Profiling out the remaining parameters and omitting terms that do not depend on \(\bm{\Phi}\) yields the following estimator of the common predictor envelope subspace:
\begin{equation}\label{eq:groupwise-predictor-envelope-objective}
\widehat{\mathcal{E}}_{\mathcal{M}_{\bm{X}}}(\mathcal{B}^{\prime})
=
\underset{\operatorname{span}(\bm{\Phi})\in\mathcal{G}(q,p)}{\operatorname{argmin}}
\left\{
\sum_{\ell=1}^{L}
f_\ell
\log\left|
\bm{\Phi}^{\top}
\bm{S}_{\bm{X}_{(\ell)}\mid\bm{Y}_{(\ell)}}
\bm{\Phi}
\right|
+
\log\left|
\bm{\Phi}^{\top}
\widehat{\bm{\Sigma}}_{\bm{X}}^{-1}
\bm{\Phi}
\right|
\right\}.
\end{equation}

Let \(\widehat{\bm{\Phi}}\in\mathbb{R}^{p\times q}\) be any semi-orthogonal basis matrix of \(\widehat{\mathcal{E}}_{\mathcal{M}_{\bm{X}}}(\mathcal{B}^{\prime})\), and let \(\widehat{\bm{\Phi}}_0\in\mathbb{R}^{p\times(p-q)}\) be an orthogonal completion of \(\widehat{\bm{\Phi}}\). Once \(\widehat{\bm{\Phi}}\) has been obtained, the remaining model parameters are estimated by
\begin{align}\label{eq:groupwise-parameter-estimators}
\widehat{\bm{\mu}}_{\bm{X}_{(\ell)}}&=\overline{\bm{X}}_{(\ell)}, \qquad \widehat{\bm{\mu}}_{(\ell)}=\overline{\bm{Y}}_{(\ell)},\\
\widehat{\bm{\eta}}_{(\ell)}&=\left(\widehat{\bm{\Phi}}^{\top}\bm{S}_{\bm{X}_{(\ell)}}\widehat{\bm{\Phi}}\right)^{-1}\widehat{\bm{\Phi}}^{\top}\bm{S}_{\bm{X}_{(\ell)}\bm{Y}_{(\ell)}},\\
\widehat{\bm{\beta}}_{(\ell)}&=\widehat{\bm{\eta}}_{(\ell)}^{\top}\widehat{\bm{\Phi}}^{\top}=\bm{S}_{\bm{Y}_{(\ell)}\bm{X}_{(\ell)}}\widehat{\bm{\Phi}}\left(\widehat{\bm{\Phi}}^{\top}\bm{S}_{\bm{X}_{(\ell)}}\widehat{\bm{\Phi}}\right)^{-1}\widehat{\bm{\Phi}}^{\top},\\
\widehat{\bm{\Delta}}_{(\ell)}&=\widehat{\bm{\Phi}}^{\top}\bm{S}_{\bm{X}_{(\ell)}}\widehat{\bm{\Phi}},\\
\widehat{\bm{\Delta}}_0&=\widehat{\bm{\Phi}}_0^{\top}\widehat{\bm{\Sigma}}_{\bm{X}}\widehat{\bm{\Phi}}_0,\\
\widehat{\bm{\Sigma}}_{\bm{X}_{(\ell)}}&=\widehat{\bm{\Phi}}\widehat{\bm{\Delta}}_{(\ell)}\widehat{\bm{\Phi}}^{\top}+\widehat{\bm{\Phi}}_0\widehat{\bm{\Delta}}_0\widehat{\bm{\Phi}}_0^{\top},\\
\widehat{\bm{\Sigma}}_{(\ell)}&=\bm{S}_{\bm{Y}_{(\ell)}}-\bm{S}_{\bm{Y}_{(\ell)}\bm{X}_{(\ell)}}\widehat{\bm{\Phi}}\left(\widehat{\bm{\Phi}}^{\top}\bm{S}_{\bm{X}_{(\ell)}}\widehat{\bm{\Phi}}\right)^{-1}\widehat{\bm{\Phi}}^{\top}\bm{S}_{\bm{X}_{(\ell)}\bm{Y}_{(\ell)}}, \qquad \ell=1,\ldots,L.
\label{eq:groupwise-parameter-estimators_end}
\end{align}
Here, \(\widehat{\bm{\Sigma}}_{(\ell)}\) is the estimator of the group-specific error covariance matrix, whereas \(\widehat{\bm{\Delta}}_0\) is obtained by projecting the pooled within-group covariance estimator onto the estimated immaterial subspace.

For comparison, the ordinary least squares estimator in group \(\ell\) is
\[
\widehat{\bm{\beta}}_{(\ell),\mathrm{OLS}}
=
\bm{S}_{\bm{Y}_{(\ell)}\bm{X}_{(\ell)}}
\bm{S}_{\bm{X}_{(\ell)}}^{-1}.
\]
The proposed estimator can therefore be expressed as
\[
\widehat{\bm{\beta}}_{(\ell)}
=
\widehat{\bm{\beta}}_{(\ell),\mathrm{OLS}}
\bm{S}_{\bm{X}_{(\ell)}}
\widehat{\bm{\Phi}}
\left(
\widehat{\bm{\Phi}}^{\top}
\bm{S}_{\bm{X}_{(\ell)}}
\widehat{\bm{\Phi}}
\right)^{-1}
\widehat{\bm{\Phi}}^{\top}.
\]
Thus, the proposed estimator restricts the group-specific regression coefficient to the estimated common predictor envelope subspace.

The objective function in \eqref{eq:groupwise-predictor-envelope-objective} depends only on the subspace spanned by \(\bm{\Phi}\), rather than on a particular choice of its orthonormal basis. Indeed, replacing \(\bm{\Phi}\) with \(\bm{\Phi}\bm{O}\), where \(\bm{O}\in\mathbb{R}^{q\times q}\) is orthogonal, leaves the objective function unchanged. Consequently, the parameter of interest is an element of \(\mathcal{G}(q,p)\).

For numerical computation, we adapt the non-Grassmann coordinate representation proposed by \citet{CookForzaniSu2016FastEnvelope} to the weighted multi-group objective in \eqref{eq:groupwise-predictor-envelope-objective}. This representation locally parameterizes a \(q\)-dimensional subspace using \(q(p-q)\) unconstrained coordinates, thereby avoiding direct constrained optimization over semi-orthogonal matrices. The resulting smooth objective function is minimized numerically with respect to these unconstrained coordinates, and an orthonormal basis \(\widehat{\bm{\Phi}}\) of the estimated subspace is obtained by orthonormalizing the resulting basis matrix. This computational reparameterization does not alter the subspace estimator defined in \eqref{eq:groupwise-predictor-envelope-objective}.
\subsection{Theoretical properties}\label{sec:theory}

In this section, we establish the asymptotic properties of the proposed estimator. We first derive the asymptotic distribution of the group-specific regression coefficient estimator and then compare its asymptotic variance with that of the estimator obtained by fitting a predictor envelope model separately to each group.

\begin{theorem}[Asymptotic distribution of the proposed estimator]
\label{thm:genv-asymptotic-distribution}
Suppose that \(L\) is fixed, \(f_\ell=n_\ell/n\in(0,1)\) is fixed with respect to \(n\) for each \(\ell\), and the envelope dimension \(q\) is known. Then, for each \(\ell=1,\ldots,L\),
\begin{equation}
\sqrt{n}\left\{\operatorname{vec}\left(\widehat{\bbeta}_{(\ell),\mathrm{genv}}\right)-\operatorname{vec}\left(\bbeta_{(\ell)}\right)\right\}\xrightarrow{d}N\left(\bm{0},\bV_{(\ell),\mathrm{genv}}\right),
\end{equation}
where
\begin{equation}
\label{eq:avar-genv}
\bV_{(\ell),\mathrm{genv}}
=
f_\ell^{-1}
\left(
\bPhi\bDelta_{(\ell)}^{-1}\bPhi^\top
\otimes
\bSigma_{(\ell)}
\right)
+
\left(
\bPhi_0\otimes\boldeta_{(\ell)}^\top
\right)
\bM^\dagger
\left(
\bPhi_0^\top\otimes\boldeta_{(\ell)}
\right),
\end{equation}
with
\begin{equation}
\label{eq:M-matrix}
\bM
=
\sum_{\ell=1}^{L}
f_{\ell}
\left\{
\bDelta_0\otimes
\boldeta_{(\ell)}
\bSigma_{(\ell)}^{-1}
\boldeta_{(\ell)}^\top
+
\bDelta_0^{-1}\otimes\bDelta_{(\ell)}
+
\bDelta_0\otimes\bDelta_{(\ell)}^{-1}
-
2\bI_{p-q}\otimes\bI_q
\right\}.
\end{equation}
\end{theorem}

\begin{theorem}[Asymptotic distribution of the separate predictor envelope estimator]
\label{thm:sep-asymptotic-distribution}
Suppose that the assumptions of Theorem~\ref{thm:genv-asymptotic-distribution} hold. Let \(\widehat{\bbeta}_{(\ell),\mathrm{sep}}\) denote the estimator of the coefficient matrix obtained by fitting a predictor envelope model separately to group \(\ell\). Then, for each \(\ell=1,\ldots,L\),
\begin{equation}
\sqrt{n}\left\{\operatorname{vec}\left(\widehat{\bbeta}_{(\ell),\mathrm{sep}}\right)-\operatorname{vec}\left(\bbeta_{(\ell)}\right)\right\}\xrightarrow{d}N\left(\bm{0},\bV_{(\ell),\mathrm{sep}}\right),
\end{equation}
where
\begin{equation}
\label{eq:avar-sep}
\bV_{(\ell),\mathrm{sep}}
=
f_\ell^{-1}
\left(
\bPhi\bDelta_{(\ell)}^{-1}\bPhi^\top
\otimes
\bSigma_{(\ell)}
\right)
+
\left(
\bPhi_0\otimes\boldeta_{(\ell)}^\top
\right)
\bm M_{(\ell)}^\dagger
\left(
\bPhi_0^\top\otimes\boldeta_{(\ell)}
\right),
\end{equation}
with
\begin{equation}
\label{eq:T-matrix}
\bm M_{(\ell)}
=
f_\ell
\left\{
\bDelta_0\otimes
\boldeta_{(\ell)}
\bSigma_{(\ell)}^{-1}
\boldeta_{(\ell)}^\top
+
\bDelta_0^{-1}\otimes\bDelta_{(\ell)}
+
\bDelta_0\otimes\bDelta_{(\ell)}^{-1}
-
2\bI_{p-q}\otimes\bI_q
\right\}.
\end{equation}
\end{theorem}
We can obtain the following corollary from Theorems \ref{thm:genv-asymptotic-distribution} and \ref{thm:sep-asymptotic-distribution}. 
\begin{corollary}[Asymptotic efficiency]
\label{cor:asymptotic-efficiency}
Suppose that the assumptions of Theorems~\ref{thm:genv-asymptotic-distribution} and~\ref{thm:sep-asymptotic-distribution} hold. Then, for each \(\ell=1,\ldots,L\),
\begin{equation}
\label{eq:efficiency-order}
\bV_{(\ell),\mathrm{sep}}
\succeq
\bV_{(\ell),\mathrm{genv}}.
\end{equation}
\end{corollary}

The first term in \eqref{eq:avar-genv} is the asymptotic variance that would remain if the common predictor envelope subspace were known. The factor \(f_\ell^{-1}\) reflects that the group-specific regression coefficient is estimated using the \(n_\ell\) observations from group \(\ell\). The second term is the additional variation attributable to estimating the common predictor envelope subspace.

The ordering in \eqref{eq:efficiency-order} shows that, in the positive semidefinite sense, the proposed estimator is at least as efficient as the estimator obtained by fitting a predictor envelope model separately to each group. This efficiency gain results from pooling information across groups when estimating the common predictor envelope subspace.

When \(L=1\), we have \(f_1=1\), and the groupwise predictor envelope model reduces to the standard predictor envelope model. In this case, \eqref{eq:avar-genv} reduces to the corresponding asymptotic variance formula given by \citet{Cook2013PLS}. Thus, Theorem~\ref{thm:genv-asymptotic-distribution} extends the asymptotic variance decomposition for the standard predictor envelope model to the multi-group setting. The pooling mechanism underlying \eqref{eq:efficiency-order} also parallels that of the groupwise response envelope model studied by \citet{park2017groupwise}.

Second, the following theorem shows the asymptotic conditional prediction errors of the proposed groupwise predictor envelope estimator and the separate predictor envelope estimator, together with their first-order comparison.

\begin{theorem}[Asymptotic conditional prediction errors]
\label{thm:prediction-error-comparison}
Suppose that the assumptions of Theorem~1 hold. Let \((\bm X_{(\ell),\mathrm{new}},\bm Y_{(\ell),\mathrm{new}})\) be a future observation from group \(\ell\), independent of the observed sample, and condition on \(\bm X_{(\ell),\mathrm{new}}=\bm x_{\mathrm{new}}\). Define the centered future predictor value as \(\widetilde{\bm x}=\bm x_{\mathrm{new}}-\bm\mu_{\bm X_{(\ell)}}\).
Assume also that, for each
\(m\in\{\mathrm{genv},\mathrm{sep}\}\), the sequence
\[
\left\{
n
\left\|
\operatorname{vec}
\left(
\widehat{\bm\beta}_{(\ell),m}
-
\bm\beta_{(\ell)}
\right)
\right\|^2
:
n\geq1
\right\}
\]
is uniformly integrable.
The conditional prediction error matrix of the proposed groupwise predictor envelope estimator satisfies
\begin{equation}
\begin{aligned}
\bm R_{(\ell),\mathrm{genv}}(\bm x_{\mathrm{new}})
={}&
\bm\Sigma_{(\ell)}
+
\frac{1}{n}f_\ell^{-1}
\left\{
1+
\widetilde{\bm x}^{\top}
\bm\Phi
\bm\Delta_{(\ell)}^{-1}
\bm\Phi^{\top}
\widetilde{\bm x}
\right\}
\bm\Sigma_{(\ell)}
\\
&+
\frac{1}{n}
\left(
\widetilde{\bm x}^{\top}\bm\Phi_0
\otimes
\bm\eta_{(\ell)}^{\top}
\right)
\bm M^{\dagger}
\left(
\bm\Phi_0^{\top}\widetilde{\bm x}
\otimes
\bm\eta_{(\ell)}
\right)
+
o\left(n^{-1}\right).
\end{aligned}
\end{equation}

The conditional prediction error matrix of the separate predictor envelope estimator satisfies
\begin{equation}
\begin{aligned}
\bm R_{(\ell),\mathrm{sep}}(\bm x_{\mathrm{new}})
={}&
\bm\Sigma_{(\ell)}
+
\frac{1}{n}f_\ell^{-1}
\left\{
1+
\widetilde{\bm x}^{\top}
\bm\Phi
\bm\Delta_{(\ell)}^{-1}
\bm\Phi^{\top}
\widetilde{\bm x}
\right\}
\bm\Sigma_{(\ell)}
\\
&+
\frac{1}{n}
\left(
\widetilde{\bm x}^{\top}\bm\Phi_0
\otimes
\bm\eta_{(\ell)}^{\top}
\right)
\bm M^{\dagger}_{(\ell)}
\left(
\bm\Phi_0^{\top}\widetilde{\bm x}
\otimes
\bm\eta_{(\ell)}
\right)
+
o\left(n^{-1}\right).
\end{aligned}
\end{equation}

Define \(\bm B_{(\ell)}(\bm x_{\mathrm{new}})=\left(\widetilde{\bm x}^{\top}\bm\Phi_0\otimes\bm\eta_{(\ell)}^{\top}\right)\left(\bm M^{\dagger}_{(\ell)}-\bm M^{\dagger}\right)\left(\bm\Phi_0^{\top}\widetilde{\bm x}\otimes\bm\eta_{(\ell)}\right)\). Then,
\[
n\left\{\bm R_{(\ell),\mathrm{sep}}(\bm x_{\mathrm{new}})-\bm R_{(\ell),\mathrm{genv}}(\bm x_{\mathrm{new}})\right\}
\longrightarrow
\bm B_{(\ell)}(\bm x_{\mathrm{new}})
\succeq
\bm 0.
\]
\end{theorem}
Therefore, the proposed groupwise predictor envelope estimator has asymptotically no larger conditional prediction error than the separate predictor envelope estimator.

\section{Numerical Experiments}\label{sec:simulation}
In this section, we evaluate the proposed method through numerical studies.
Section 4.1 presents Monte Carlo simulations for assessing its finite-sample performance.
Section 4.2 applies the proposed method to the Wine data set.
\subsection{Monte Carlo Simulation}
\subsubsection{Comparison of estimation and prediction accuracy}\label{sec:normal_sim}

We first compared the estimation and prediction accuracy of the proposed groupwise predictor envelope estimator with ordinary least squares and the separate predictor envelope estimator.

We considered two groups, \(L=2\). For each group \(\ell=1,2\), the predictor vector \(\bm{X}_{(\ell)j}\in\mathbb{R}^{20}\) was generated independently from
\[
\bm{X}_{(\ell)j}
\sim
N_{20}\!\left(\bm{0},\bm{\Sigma}_{X(\ell)}\right),
\qquad
j=1,\ldots,n_\ell.
\]
The response vector \(\bm{Y}_{(\ell)j}\in\mathbb{R}^{10}\) was generated according to
\[
\bm{Y}_{(\ell)j}
=
\bm{\eta}_{(\ell)}^{\top}
\bm{\Phi}^{\top}
\bm{X}_{(\ell)j}
+
\bm{\varepsilon}_{(\ell)j},
\qquad
\bm{\varepsilon}_{(\ell)j}
\sim
N_{10}\!\left(\bm{0},5\ell\bm{I}_{10}\right).
\]
The predictor covariance matrix for group \(\ell\) was specified as
\[
\bm{\Sigma}_{X(\ell)}
=
\bm{\Phi}
\bm{\Delta}_{(\ell)}
\bm{\Phi}^{\top}
+
\bm{\Phi}_0
\bm{\Delta}_0
\bm{\Phi}_0^{\top},
\]
where
\[
\bm{\Delta}_{(1)}=
\begin{pmatrix}
    10 & 0 \\
    0 & 2
\end{pmatrix}, \quad \bm{\Delta}_{(2)}=
\begin{pmatrix}
    60 & 0 \\
    0 & 12
\end{pmatrix}, 
\]
and the eigenvalues of $\bm{\Delta}_0$ were specified as
\[
d_j
=
0.5\left(\frac{0.05}{0.5}\right)^{(j-1)/17},
\qquad j=1,\ldots,18,
\]
and $\bm{\Delta}_0=\operatorname{diag}(d_1,\ldots,d_{18})$.
Thus, the two groups shared the same predictor envelope subspace but had different amounts of material predictor variation.
The dimension of the predictor envelope subspace was set to \(q=2\). The orthogonal matrix \((\bm{\Phi},\bm{\Phi}_0)\) was constructed by orthogonalizing a \(20\times20\) random matrix whose entries were independently generated from the standard normal distribution. Each element of the group-specific coefficient matrix \(\bm{\eta}_{(\ell)}\in\mathbb{R}^{2\times10}\) was independently generated from a normal distribution with mean zero and variance \(100\).
The total sample size was set to
\[
n\in\{100,300,1000,2000\},
\]
and the group sample sizes were given by
\[
n_1=0.4n,
\qquad
n_2=0.6n.
\]
For each sample-size setting, 100 Monte Carlo replications were conducted.

We compared the following three estimators: ordinary least squares (OLS), the separate predictor envelope estimator, which fits a predictor envelope model independently within each group, and the proposed groupwise predictor envelope estimator, which estimates the common predictor envelope subspace jointly across groups.

Estimation accuracy was evaluated using the empirical standard deviations of the estimated regression coefficients. For each coefficient, its empirical standard deviation was calculated over the 100 Monte Carlo replications, and these standard deviations were then averaged over all coefficients within each group. Prediction accuracy was evaluated using the mean squared prediction error on an independently generated test sample.

\paragraph{Estimation accuracy.}
\begin{table}[!htbp]
    \centering
    \caption{
        Average empirical standard deviations of the estimated regression coefficients over 100 Monte Carlo replications.
        Lower values indicate more accurate estimation.
    }
    \label{tab:coefficient_se}
    \small
    \begin{tabular}{cccccc}
        \toprule
        Group & \(n\) & \(n_{\ell}\) & OLS & Separate PE & Proposed method \\
        \midrule
        1 & 100  & 40   & 1.4118 & 0.3121 & \textbf{0.1307} \\
        1 & 300  & 120  & 0.5943 & 0.1459 & \textbf{0.0692} \\
        1 & 1000 & 400  & 0.3091 & 0.0762 & \textbf{0.0363} \\
        1 & 2000 & 800  & 0.2141 & 0.0533 & \textbf{0.0266} \\
        \midrule
        2 & 100  & 60   & 1.3607 & 0.1459 & \textbf{0.1288} \\
        2 & 300  & 180  & 0.6692 & 0.0768 & \textbf{0.0688} \\
        2 & 1000 & 600  & 0.3542 & 0.0394 & \textbf{0.0352} \\
        2 & 2000 & 1200 & 0.2491 & 0.0292 & \textbf{0.0259} \\
        \bottomrule
    \end{tabular}
\end{table}
Table~\ref{tab:coefficient_se} shows the average empirical standard deviations of the estimated regression coefficients for the two groups.
The proposed method achieved the smallest empirical standard errors for every sample size in both groups. For example, when \(n=100\), the average empirical standard deviations of the proposed estimator were \(0.1307\) and \(0.1288\) for Groups 1 and 2, respectively. The corresponding values were \(0.3121\) and \(0.1459\) for the separate predictor envelope estimator and \(1.4118\) and \(1.3607\) for OLS.

These results indicate that jointly estimating the common predictor envelope subspace improves the estimation efficiency of the group-specific regression coefficients. The separate predictor envelope estimator also substantially improved upon OLS, but its standard deviations were larger than those of the proposed estimator because the common envelope subspace was estimated separately using only the observations within each group.

\paragraph{Prediction accuracy.}

Table~\ref{tab:test_mse_with_se} shows the distributions of the test mean squared prediction errors over the 100 Monte Carlo replications.
\begin{table}[htbp]
    \centering
    \caption{
        Mean test prediction errors with Monte Carlo standard errors
        over 100 Monte Carlo replications.
        Monte Carlo standard errors are reported in parentheses.
        Lower values indicate more accurate prediction.
    }
    \label{tab:test_mse_with_se}
    \small
    \begin{tabular}{cccrrr}
        \toprule
        Group & \(n\) & \(n_{\ell}\)
        & OLS & Separate PE & Proposed method \\
        \midrule
        1 & 100  & 40
          & 10.835 (0.122)
          & 5.853 (0.043)
          & \textbf{5.470 (0.039)} \\
        1 & 300  & 120
          & 6.033 (0.030)
          & 5.195 (0.024)
          & \textbf{5.130 (0.023)} \\
        1 & 1000 & 400
          & 5.296 (0.013)
          & 5.077 (0.012)
          & \textbf{5.059 (0.012)} \\
        1 & 2000 & 800
          & 5.145 (0.008)
          & 5.040 (0.007)
          & \textbf{5.031 (0.007)} \\
        \midrule
        2 & 100  & 60
          & 15.344 (0.107)
          & 10.564 (0.071)
          & \textbf{10.511 (0.068)} \\
        2 & 300  & 180
          & 11.342 (0.040)
          & 10.231 (0.034)
          & \textbf{10.221 (0.034)} \\
        2 & 1000 & 600
          & 10.356 (0.020)
          & 10.050 (0.019)
          & \textbf{10.048 (0.019)} \\
        2 & 2000 & 1200
          & 10.160 (0.013)
          & 10.007 (0.012)
          & \textbf{10.005 (0.012)} \\
        \bottomrule
    \end{tabular}
\end{table}

The proposed method had the smallest mean prediction error for both groups at every sample size. The improvement was particularly pronounced for Group 1 when the sample size was small. For example, when \(n=100\), the mean test prediction errors for Group 1 were \(5.470\), \(5.853\), and \(10.835\) for the proposed estimator, the separate predictor envelope estimator, and OLS, respectively.

For Group 2, the proposed estimator also produced smaller prediction errors than the separate predictor envelope estimator, although the difference between the two envelope-based estimators was relatively small. When \(n=100\), their mean test prediction errors were \(10.511\) and \(10.564\), respectively, and when \(n=2000\), they were \(10.005\) and \(10.007\), respectively.

OLS consistently produced larger prediction errors than the two envelope-based estimators. As the sample size increased, the prediction errors of the envelope-based estimators approached the irreducible error levels of \(5\) and \(10\) for Groups 1 and 2, respectively.

\subsubsection{Simulation with unequal group sample sizes}

We considered two groups, \(L=2\), with \(p=12\),
\(r=4\), and \(q=2\).
For each group \(\ell=1,2\), the predictors were independently generated as
\[
\bm{X}_{(\ell)j}
\sim
N_{12}\!\left(
    \bm{0},
    \bm{\Sigma}_{X_{(\ell)}}
\right),
\qquad
j=1,\ldots,n_\ell,
\]
where
\[
\bm{\Sigma}_{X(\ell)}
=
\bm{\Phi}
\bm{\Delta}_{(\ell)}
\bm{\Phi}^{\top}
+
\bm{\Phi}_0
\bm{\Delta}_0
\bm{\Phi}_0^{\top}.
\]
Here, \(\bm{\Phi}\in\mathbb{R}^{12\times2}\) represents the common
material subspace, and
\(\bm{\Phi}_0\in\mathbb{R}^{12\times10}\) is its orthogonal complement.
The orthogonal matrix $(\bm{\Phi},\bm{\Phi}_0)$ was obtained by applying a QR decomposition to a
\(12\times12\) random matrix whose entries were independently generated from the standard normal distribution.
The material covariance matrices were specified as
\[
\bm{\Delta}_{(1)}
=
\begin{pmatrix}
4.5 & 0 \\
0 & 9
\end{pmatrix},
\qquad
\bm{\Delta}_{(2)}
=
\begin{pmatrix}
7.5 & 0 \\
0 & 15
\end{pmatrix},
\]
whereas the covariance matrix of the immaterial predictor component was
\[
\bm{\Delta}_0=10\bm{I}_{10}.
\]
Thus, the two groups shared the same material subspace but had different
variability within that subspace.
The response vectors were generated according to
\[
\bm{Y}_{(\ell)j}
=
\bm{\beta}_{(\ell)}
\bm{X}_{(\ell)j}
+
\bm{\varepsilon}_{(\ell)j},
\]
where
\[
\bm{\beta}_{(\ell)}
=
\bm{\eta}_{(\ell)}^{\top}
\bm{\Phi}^{\top}.
\]
Each element of
\(\bm{\eta}_{(\ell)}\in\mathbb{R}^{2\times4}\)
was independently generated from \(N(0,1.25^2)\).
The error vectors were generated independently of the predictors as
\[
\bm{\varepsilon}_{(\ell)j}
\sim
N_4\!\left(
    \bm{0},
    \bm{\Sigma}_{\varepsilon(\ell)}
\right),
\]
with
\[
\bm{\Sigma}_{\varepsilon(1)}
=
\operatorname{diag}(0.72,0.90,1.08,1.26)
\]
and
\[
\bm{\Sigma}_{\varepsilon(2)}
=
\operatorname{diag}(0.88,1.10,1.32,1.54).
\]
All population parameters were generated once and held fixed across
Monte Carlo replications.

To investigate the effect of highly unequal group sample sizes, the
sample size of Group 1 was fixed at
\(
n_1=30,
\)
whereas the sample size of Group 2 was varied over
\(
n_2\in\{50,150,300,600\}.
\)
Consequently, the four sample size configurations were
\[
(n_1,n_2)
=
(30,50),\ (30,150),\ (30,300),\ (30,600).
\]
Each configuration was replicated 100 times.

We compared ordinary least squares (OLS), the separate predictor
envelope estimator, and the proposed method. The true envelope dimension
\(q=2\) was supplied to both envelope-based estimators.
As in Section~\ref{sec:normal_sim}, we evaluated the methods in terms of
the standard errors of the estimated regression coefficients and the test
prediction errors.

\begin{figure}[!htbp] 
    \centering
    \includegraphics[width=\linewidth]{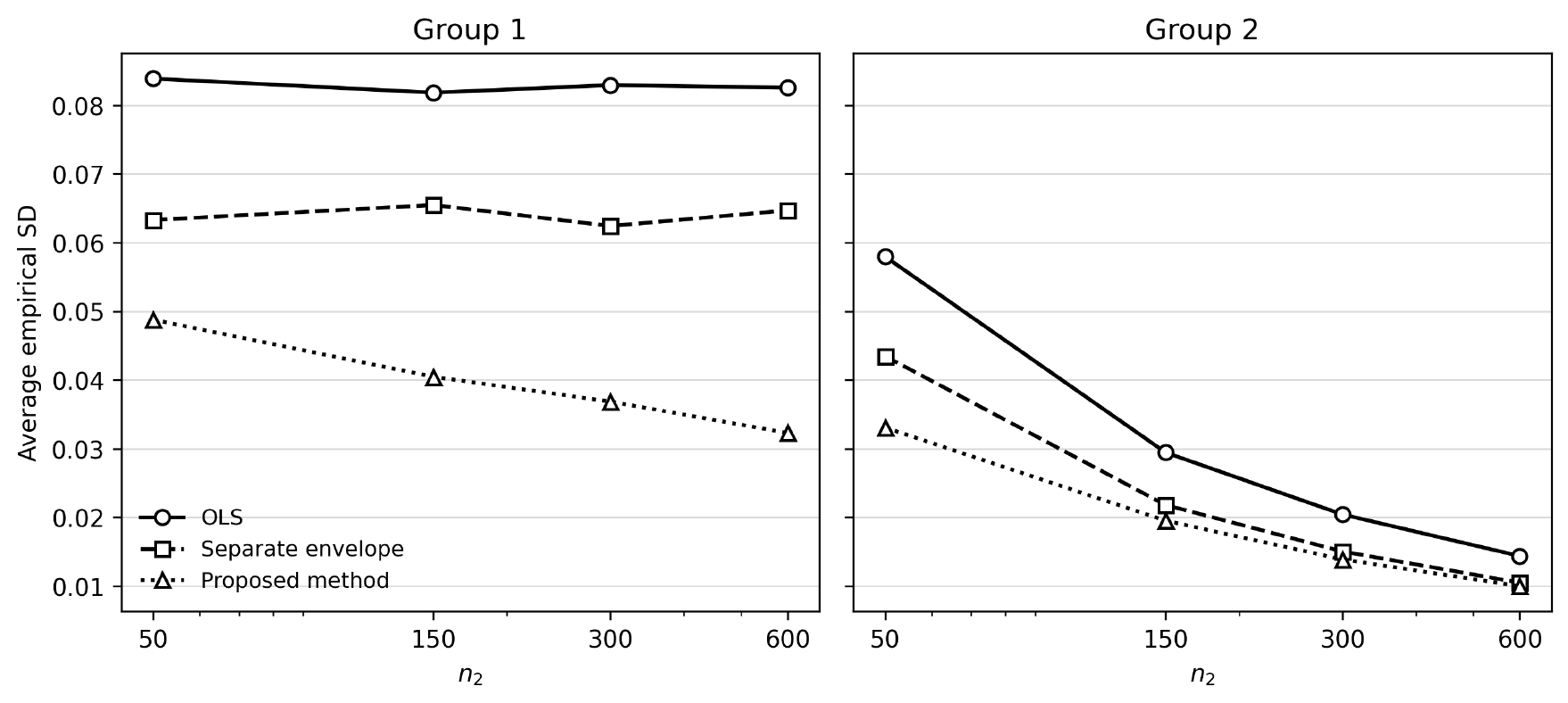}
    \caption{Average empirical standard deviations of the estimated regression coefficients over 100 Monte Carlo replications under unequal group sample sizes. The left and right panels correspond to Groups 1 and 2, respectively. Lower values indicate greater estimation efficiency.}
    \label{fig:unequal_sd}
\end{figure}

Figure~\ref{fig:unequal_sd} shows the average empirical standard deviations
of the estimated regression coefficients. The proposed method achieved the
smallest value in both groups for every sample-size configuration. In Group 1,
although the sample size was fixed at \(n_1=30\), the average empirical
standard deviation of the proposed estimator decreased monotonically from
\(0.0488\) at \(n_2=50\) to \(0.0323\) at \(n_2=600\). In contrast, the
corresponding values for OLS and the separate predictor envelope estimator
remained nearly constant as \(n_2\) increased. In Group 2, the empirical
standard deviations decreased with \(n_2\) for all three methods, with the
proposed method consistently attaining the smallest values.

\begin{figure}[!htbp] 
    \centering
    \includegraphics[width=\linewidth]{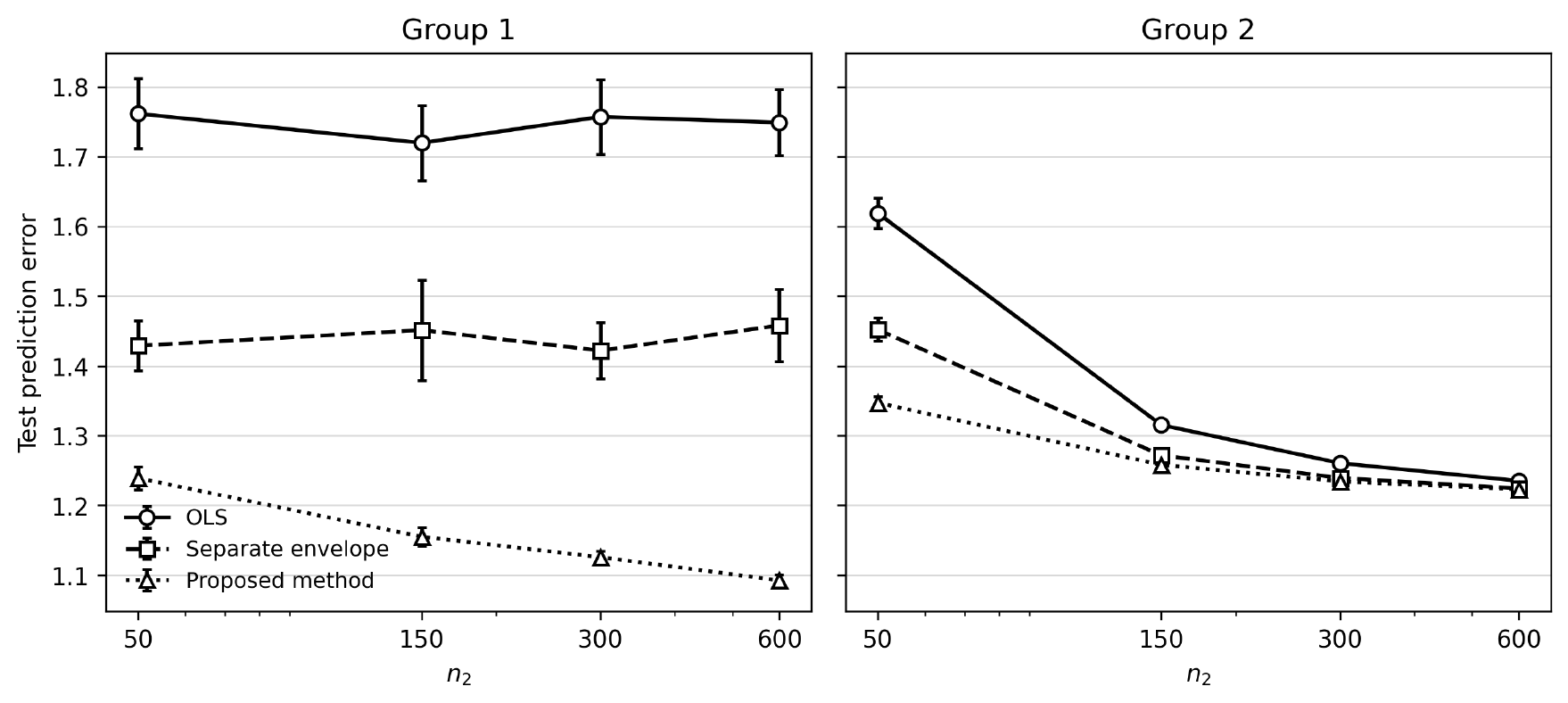}
    \caption{Average test prediction errors over 100 Monte Carlo replications under unequal group sample sizes. The left and right panels correspond to Groups 1 and 2, respectively. Lower values indicate better predictive performance.}
    \label{fig:unequal_mse}
\end{figure}

Figure~\ref{fig:unequal_mse} presents the test prediction errors. The proposed
method yielded the smallest prediction error in both groups for all
sample-size configurations. In Group 1, its prediction error decreased
steadily from \(1.2390\) at \(n_2=50\) to \(1.0926\) at \(n_2=600\), even
though \(n_1\) remained fixed. By contrast, the prediction errors of OLS and
the separate predictor envelope estimator showed no systematic improvement
as \(n_2\) increased. In Group 2, the prediction errors decreased with
increasing \(n_2\) for all methods, and the proposed method consistently
provided the best predictive performance.


These results highlight the importance of sharing information across groups
when the group sample sizes are highly unbalanced. The OLS estimator and the
separate predictor envelope estimator for Group 1 are based only on the
\(n_1=30\) observations from that group and therefore cannot benefit from the
increase in the sample size of Group 2. In contrast, the proposed method
jointly estimates the common envelope subspace using observations from both
groups. Consequently, the more accurate estimation of the common subspace
resulting from a larger \(n_2\) improves both coefficient estimation and
prediction in Group 1. This demonstrates that borrowing information through
a common envelope structure can be particularly beneficial for a group with
a small sample size.


\subsection{Application}

We applied the proposed method to the Wine data set as an example involving
multiple groups. The data set consists of three groups, denoted by Groups A, B,
and C, with sample sizes 59, 71, and 48, respectively. We treated Color
intensity, Hue, and OD280/OD315 of diluted wines as three response variables
and used the remaining ten chemical measurements as predictors. Thus, the
analysis was conducted under a multivariate linear regression setting with
three groups.

Both the predictors and responses were standardized separately within each
group. Consequently, the common predictor envelope subspace for the proposed
method is defined in the predictor coordinates obtained after standardization
within each group. In the prediction assessment, the standardization
parameters were estimated using only the training observations and then
applied to the corresponding test observations.

We compared the proposed groupwise predictor envelope estimator with ordinary
least squares (OLS) and the separate envelope estimator, which fits a predictor
envelope model independently within each group. The envelope dimensions were
selected by BIC. Using the full data, the selected common dimension was $q=3$
for the proposed method, whereas the selected dimensions for the separate
envelope estimator were $q=3$, $4$, and $2$ for Groups A, B, and C,
respectively. Standard errors and prediction errors were used as the evaluation
criteria.

\begin{table}[t]
    \centering
    \caption{Standard errors and prediction errors for the Wine data. Values in parentheses are standard deviations of the prediction errors across the repeated cross-validation runs. Smaller values are better.}
    \label{tab:wine_results}
    \begin{tabular}{llcc}
    \toprule
    Group & Method & Standard error & Prediction error \\
    \midrule
    \multirow{3}{*}{A}
    & OLS               & 0.067          & \textbf{0.770} (0.049) \\
    & Separate envelope & 0.064          & 0.820 (0.087) \\
    & Proposed          & \textbf{0.055} & 0.798 (0.051) \\
    \midrule
    \multirow{3}{*}{B}
    & OLS               & 0.077          & \textbf{0.838} (0.042) \\
    & Separate envelope & 0.088          & 0.873 (0.067) \\
    & Proposed          & \textbf{0.076} & 0.887 (0.034) \\
    \midrule
    \multirow{3}{*}{C}
    & OLS               & 0.123          & 4.494 (0.466) \\
    & Separate envelope & 0.091          & 4.653 (0.366) \\
    & Proposed          & \textbf{0.072} & \textbf{4.031} (0.286) \\
    \bottomrule
    \end{tabular}
\end{table}

Table~\ref{tab:wine_results} shows that the proposed method yielded the
smallest median coefficient standard error in all three groups. Relative to
OLS, the median standard error was reduced from 0.067 to 0.055 in Group A,
from 0.077 to 0.076 in Group B, and from 0.123 to 0.072 in Group C. These
results indicate that the proposed method generally improved the stability of
the estimated regression coefficients. Across all groups, 67 of the 90
standard errors for the individual coefficients were smaller for the proposed
method than for OLS.

The prediction results were not uniform across groups. OLS gave the smallest
prediction errors for Groups A and B, whereas the proposed method gave the
smallest prediction error for Group C. When all observations were combined,
the prediction errors were 1.801 for OLS, 1.875 for the separate envelope
estimator, and 1.705 for the proposed method. Thus, the proposed method had the
smallest overall prediction error, although its groupwise advantage was mainly
observed in Group C. Overall, the Wine data analysis indicates that the
proposed method improved the stability of the coefficient estimates in all
three groups and achieved favorable prediction performance, particularly for
Group C.

\section{Conclusion}\label{sec:conclusion}

This paper introduced a groupwise predictor envelope model for multivariate
linear regression with observations from several groups. The model estimates a
common predictor envelope subspace using all groups, while allowing the
regression coefficients, the predictor covariance within this subspace, and
the response error covariance matrices to vary across groups. We derived an
objective function for estimating the common subspace and obtained the
resulting regression estimators. We also established asymptotic normality and
derived explicit asymptotic variance matrices for the regression coefficient
estimators. The proposed estimator has no larger asymptotic variance, in the positive
semidefinite order, than the estimator obtained by fitting a predictor envelope
model separately to each group. In addition, we derived the conditional
prediction errors through order $n^{-1}$ and showed that the proposed estimator
has no larger conditional prediction error than the separate estimator to this
order.

The simulation results supported these theoretical findings. The proposed
method improved the estimation of the regression coefficients and prediction
when compared with OLS and the separate predictor envelope estimator. The
improvement was particularly clear when the group sample sizes were unequal:
increasing the sample size of the larger group improved estimation and
prediction for the smaller group because the common subspace was estimated
using observations from both groups. In the Wine data analysis, the proposed
method had the smallest median standard error in all three groups and the
smallest overall prediction error. However, its prediction error was not the
smallest in every group, indicating that the benefit in finite samples can
differ across groups.

The proposed model assumes a common predictor envelope subspace and a common
covariance for the predictor variation orthogonal to that subspace. When these
assumptions do not adequately describe the data, combining information across
groups may provide less benefit. Further work may therefore consider methods
for assessing these assumptions and models that allow part of the relevant
predictor envelope subspace to differ across groups. It would also be useful to
develop asymptotic results that account for estimating the envelope dimension,
which is treated as known in the present theory but selected from the data in
practice.

\section*{Acknowledgements}
A. O. was supported by JSPS KAKENHI Grant Number JP25K24377.
S. K. was supported by JSPS KAKENHI Grant Numbers JP23K11008,  JP23H00809,  and JP23H03352. 
\bibliographystyle{apalike}
\bibliography{reference}

\clearpage
\appendix
\section*{Appendix}
\addcontentsline{toc}{section}{Appendix}


\section{Derivation of the groupwise predictor envelope estimator}
\label{app:estimation}

This section derives the objective function in
\eqref{eq:groupwise-predictor-envelope-objective} and the estimators in
\eqref{eq:groupwise-parameter-estimators}--\eqref{eq:groupwise-parameter-estimators_end}, following the same argument
as in Section~\ref{sec:predictor-envelope}.  Recall that
$f_\ell=n_\ell/n$ and that all sample covariance matrices are computed
after centering within each group.

For
\[
\bm C_{(\ell)j}
=
\begin{pmatrix}
\bm X_{(\ell)j}\\
\bm Y_{(\ell)j}
\end{pmatrix},
\]
the sample covariance matrix is
\begin{equation}
\label{app:eq:sample-joint-covariance}
\bm S_{\bm C_{(\ell)}}
=
\begin{pmatrix}
\bm S_{\bm X_{(\ell)}}
&
\bm S_{\bm X_{(\ell)}\bm Y_{(\ell)}}
\\
\bm S_{\bm Y_{(\ell)}\bm X_{(\ell)}}
&
\bm S_{\bm Y_{(\ell)}}
\end{pmatrix}.
\end{equation}
As in Section~\ref{sec:estimation}, the model parameters are estimated
by minimizing the weighted discrepancy function
\begin{equation}
\label{app:eq:discrepancy-restated}
F\left(
\left\{\bm S_{\bm C_{(\ell)}}\right\}_{\ell=1}^{L},
\left\{\bm\Sigma_{\bm C_{(\ell)}}\right\}_{\ell=1}^{L}
\right)
=
\sum_{\ell=1}^{L}f_\ell
\left[
\log\left|\bm\Sigma_{\bm C_{(\ell)}}\right|
+
\operatorname{tr}\left(
\bm S_{\bm C_{(\ell)}}
\bm\Sigma_{\bm C_{(\ell)}}^{-1}
\right)
\right].
\end{equation}
Up to an additive constant and a positive multiplicative factor, this
is the negative Gaussian log-likelihood per observation.

\paragraph{Transformation of the joint covariance matrix.}
Under the groupwise multivariate linear regression model,
\begin{equation}
\label{app:eq:joint-covariance}
\bm\Sigma_{\bm C_{(\ell)}}
=
\begin{pmatrix}
\bm\Sigma_{\bm X_{(\ell)}}
&
\bm\Sigma_{\bm X_{(\ell)}}\bm\beta_{(\ell)}^{\top}
\\
\bm\beta_{(\ell)}\bm\Sigma_{\bm X_{(\ell)}}
&
\bm\Sigma_{\bm Y_{(\ell)}}
\end{pmatrix},
\end{equation}
where
\begin{equation}
\label{app:eq:marginal-Y-covariance}
\bm\Sigma_{\bm Y_{(\ell)}}
=
\bm\beta_{(\ell)}
\bm\Sigma_{\bm X_{(\ell)}}
\bm\beta_{(\ell)}^{\top}
+
\bm\Sigma_{(\ell)}
=
\bm\eta_{(\ell)}^{\top}
\bm\Delta_{(\ell)}
\bm\eta_{(\ell)}
+
\bm\Sigma_{(\ell)}.
\end{equation}
Using
\[
\bm\beta_{(\ell)}^{\top}
=
\bm\Phi\bm\eta_{(\ell)}
\quad\text{and}\quad
\bm\Sigma_{\bm X_{(\ell)}}
=
\bm\Phi\bm\Delta_{(\ell)}\bm\Phi^{\top}
+
\bm\Phi_0\bm\Delta_0\bm\Phi_0^{\top},
\]
we can write \eqref{app:eq:joint-covariance} as
\begin{align}
\bm\Sigma_{\bm C_{(\ell)}}
&=
\begin{pmatrix}
\bm\Phi\bm\Delta_{(\ell)}\bm\Phi^{\top}
+
\bm\Phi_0\bm\Delta_0\bm\Phi_0^{\top}
&
\bm\Phi\bm\Delta_{(\ell)}\bm\eta_{(\ell)}
\\
\bm\eta_{(\ell)}^{\top}\bm\Delta_{(\ell)}\bm\Phi^{\top}
&
\bm\Sigma_{\bm Y_{(\ell)}}
\end{pmatrix}
\nonumber\\
&=
\bm O
\bm\Sigma_{\bm O^{\top}\bm C_{(\ell)}}
\bm O^{\top},
\label{app:eq:joint-covariance-factorization}
\end{align}
where
\begin{equation}
\label{app:eq:O-and-transformed-covariance}
\bm O
=
\begin{pmatrix}
\bm\Phi & \bm\Phi_0 & \bm 0\\
\bm 0 & \bm 0 & \bm I_r
\end{pmatrix},
\qquad
\bm\Sigma_{\bm O^{\top}\bm C_{(\ell)}}
=
\begin{pmatrix}
\bm\Delta_{(\ell)}
& \bm 0
& \bm\Delta_{(\ell)}\bm\eta_{(\ell)}
\\
\bm 0
& \bm\Delta_0
& \bm 0
\\
\bm\eta_{(\ell)}^{\top}\bm\Delta_{(\ell)}
& \bm 0
& \bm\Sigma_{\bm Y_{(\ell)}}
\end{pmatrix}.
\end{equation}
The matrix $\bm O$ is orthogonal because
$(\bm\Phi,\bm\Phi_0)$ is orthogonal.

To place the material predictor component and the response next to
each other, define the permutation matrix
\[
\bm P
=
\begin{pmatrix}
\bm I_q & \bm 0 & \bm 0\\
\bm 0 & \bm 0 & \bm I_{p-q}\\
\bm 0 & \bm I_r & \bm 0
\end{pmatrix}.
\]
Then
\begin{equation}
\label{app:eq:permuted-joint-covariance}
\bm P^{\top}
\bm\Sigma_{\bm O^{\top}\bm C_{(\ell)}}
\bm P
=
\begin{pmatrix}
\bm A_{(\ell)} & \bm 0\\
\bm 0 & \bm\Delta_0
\end{pmatrix},
\qquad
\bm A_{(\ell)}
=
\begin{pmatrix}
\bm\Delta_{(\ell)}
&
\bm\Delta_{(\ell)}\bm\eta_{(\ell)}
\\
\bm\eta_{(\ell)}^{\top}\bm\Delta_{(\ell)}
&
\bm\Sigma_{\bm Y_{(\ell)}}
\end{pmatrix}.
\end{equation}

\paragraph{Separation of the discrepancy function.}
Since both $\bm O$ and $\bm P$ are orthogonal,
\begin{align}
\log\left|\bm\Sigma_{\bm C_{(\ell)}}\right|
&=
\log\left|
\bm\Sigma_{\bm O^{\top}\bm C_{(\ell)}}
\right|
\nonumber\\
&=
\log\left|
\bm P^{\top}
\bm\Sigma_{\bm O^{\top}\bm C_{(\ell)}}
\bm P
\right|
\nonumber\\
&=
\log|\bm A_{(\ell)}|
+
\log|\bm\Delta_0|.
\label{app:eq:logdet-separation}
\end{align}

For the trace term, let
\[
\bm Z_{(\ell)j}
=
\begin{pmatrix}
\bm\Phi^{\top}\bm X_{(\ell)j}\\
\bm Y_{(\ell)j}
\end{pmatrix}.
\]
Its sample covariance matrix is
\begin{equation}
\label{app:eq:sample-Z-covariance}
\bm S_{\bm Z_{(\ell)}}
=
\begin{pmatrix}
\bm\Phi^{\top}
\bm S_{\bm X_{(\ell)}}
\bm\Phi
&
\bm\Phi^{\top}
\bm S_{\bm X_{(\ell)}\bm Y_{(\ell)}}
\\
\bm S_{\bm Y_{(\ell)}\bm X_{(\ell)}}
\bm\Phi
&
\bm S_{\bm Y_{(\ell)}}
\end{pmatrix}.
\end{equation}
Let
$\bm S_{\bm O^{\top}\bm C_{(\ell)}}
=\bm O^{\top}\bm S_{\bm C_{(\ell)}}\bm O$.
Using the same orthogonal transformation and permutation in the trace
term gives
\begin{align}
&\operatorname{tr}\left(
\bm S_{\bm C_{(\ell)}}
\bm\Sigma_{\bm C_{(\ell)}}^{-1}
\right)
\nonumber\\
&\quad=
\operatorname{tr}\left(
\bm S_{\bm O^{\top}\bm C_{(\ell)}}
\bm\Sigma_{\bm O^{\top}\bm C_{(\ell)}}^{-1}
\right)
\nonumber\\
&\quad=
\operatorname{tr}\left[
\left(
\bm P^{\top}
\bm S_{\bm O^{\top}\bm C_{(\ell)}}
\bm P
\right)
\left(
\bm P^{\top}
\bm\Sigma_{\bm O^{\top}\bm C_{(\ell)}}
\bm P
\right)^{-1}
\right]
\nonumber\\
&\quad=
\operatorname{tr}\left\{
\bm S_{\bm Z_{(\ell)}}
\bm A_{(\ell)}^{-1}
\right\}
+
\operatorname{tr}\left\{
\bm\Phi_0^{\top}
\bm S_{\bm X_{(\ell)}}
\bm\Phi_0
\bm\Delta_0^{-1}
\right\}.
\label{app:eq:trace-separation}
\end{align}

Substituting \eqref{app:eq:logdet-separation} and
\eqref{app:eq:trace-separation} into
\eqref{app:eq:discrepancy-restated}, and using
\[
\widehat{\bm\Sigma}_{\bm X}
=
\sum_{\ell=1}^{L}f_\ell
\bm S_{\bm X_{(\ell)}},
\qquad
\sum_{\ell=1}^{L}f_\ell=1,
\]
we obtain
\begin{align}
\label{app:eq:separated-discrepancy}
F\left(
\left\{\bm S_{\bm C_{(\ell)}}\right\}_{\ell=1}^{L},
\left\{\bm\Sigma_{\bm C_{(\ell)}}\right\}_{\ell=1}^{L}
\right)
={}&
\sum_{\ell=1}^{L}f_\ell
\left[
\log|\bm A_{(\ell)}|
+
\operatorname{tr}\left\{
\bm S_{\bm Z_{(\ell)}}
\bm A_{(\ell)}^{-1}
\right\}
\right]
\nonumber\\
&+
\log|\bm\Delta_0|
+
\operatorname{tr}\left\{
\bm\Phi_0^{\top}
\widehat{\bm\Sigma}_{\bm X}
\bm\Phi_0
\bm\Delta_0^{-1}
\right\}.
\end{align}
For fixed $\bm\Phi$, the $\ell$th term in the sum depends only on
$\bm A_{(\ell)}$, whereas the last two terms depend only on
$\bm\Delta_0$.  Minimizing them separately gives
\begin{equation}
\label{app:eq:profiled-blocks}
\widehat{\bm A}_{(\ell)}
=
\bm S_{\bm Z_{(\ell)}},
\qquad
\widehat{\bm\Delta}_0
=
\bm\Phi_0^{\top}
\widehat{\bm\Sigma}_{\bm X}
\bm\Phi_0.
\end{equation}
Writing the first equality in \eqref{app:eq:profiled-blocks} explicitly,
\begin{equation}
\label{app:eq:block-equation}
\begin{pmatrix}
\widehat{\bm\Delta}_{(\ell)}
&
\widehat{\bm\Delta}_{(\ell)}\widehat{\bm\eta}_{(\ell)}
\\
\widehat{\bm\eta}_{(\ell)}^{\top}\widehat{\bm\Delta}_{(\ell)}
&
\widehat{\bm\Sigma}_{\bm Y_{(\ell)}}
\end{pmatrix}
=
\begin{pmatrix}
\bm\Phi^{\top}\bm S_{\bm X_{(\ell)}}\bm\Phi
&
\bm\Phi^{\top}\bm S_{\bm X_{(\ell)}\bm Y_{(\ell)}}
\\
\bm S_{\bm Y_{(\ell)}\bm X_{(\ell)}}\bm\Phi
&
\bm S_{\bm Y_{(\ell)}}
\end{pmatrix}.
\end{equation}
It follows that
\begin{equation}
\label{app:eq:eta-hat}
\widehat{\bm\eta}_{(\ell)}
=
\left(\bm\Phi^{\top}\bm S_{\bm X_{(\ell)}}\bm\Phi\right)^{-1}
\bm\Phi^{\top}\bm S_{\bm X_{(\ell)}\bm Y_{(\ell)}}.
\end{equation}
Solving \eqref{app:eq:marginal-Y-covariance} for the group-specific
error covariance matrix and using \eqref{app:eq:block-equation} and
\eqref{app:eq:eta-hat} gives
\begin{align}
\widehat{\bm\Sigma}_{(\ell)}
={}&
\widehat{\bm\Sigma}_{\bm Y_{(\ell)}}
-
\widehat{\bm\eta}_{(\ell)}^{\top}
\widehat{\bm\Delta}_{(\ell)}
\widehat{\bm\eta}_{(\ell)}
\nonumber\\
={}&
\bm S_{\bm Y_{(\ell)}}
-
\bm S_{\bm Y_{(\ell)}\bm X_{(\ell)}}
\bm\Phi
\left(
\bm\Phi^{\top}
\bm S_{\bm X_{(\ell)}}
\bm\Phi
\right)^{-1}
\bm\Phi^{\top}
\bm S_{\bm X_{(\ell)}\bm Y_{(\ell)}}.
\label{app:eq:error-covariance-hat}
\end{align}
The predictor covariance estimator for fixed $\bm\Phi$ is therefore
\begin{align}
\widehat{\bm\Sigma}_{\bm X_{(\ell)}}
={}&
\bm\Phi
\widehat{\bm\Delta}_{(\ell)}
\bm\Phi^{\top}
+
\bm\Phi_0
\widehat{\bm\Delta}_0
\bm\Phi_0^{\top}.
\label{app:eq:predictor-covariance-hat}
\end{align}

\paragraph{Profile objective function for $\bm\Phi$.}
Substituting \eqref{app:eq:profiled-blocks} into
\eqref{app:eq:separated-discrepancy} gives
\begin{align}
F\left(
\left\{\bm S_{\bm C_{(\ell)}}\right\}_{\ell=1}^{L},
\left\{\bm\Sigma_{\bm C_{(\ell)}}\right\}_{\ell=1}^{L}
\right)
={}&
\sum_{\ell=1}^{L}f_\ell
\log
\left|
\bm S_{\bm Z_{(\ell)}}
\right|
+
\log
\left|
\bm\Phi_0^{\top}
\widehat{\bm\Sigma}_{\bm X}
\bm\Phi_0
\right|
+
p+r.
\label{app:eq:first-profile-objective}
\end{align}

It remains to rewrite the two determinant terms.  First, the Schur
complement formula gives
\begin{align}
\left|
\bm S_{\bm Z_{(\ell)}}
\right|
&=
\left|
\bm S_{\bm Y_{(\ell)}}
\right|
\left|
\bm\Phi^{\top}
\left(
\bm S_{\bm X_{(\ell)}}
-
\bm S_{\bm X_{(\ell)}\bm Y_{(\ell)}}
\bm S_{\bm Y_{(\ell)}}^{-1}
\bm S_{\bm Y_{(\ell)}\bm X_{(\ell)}}
\right)
\bm\Phi
\right|
\nonumber\\
&=
\left|
\bm S_{\bm Y_{(\ell)}}
\right|
\left|
\bm\Phi^{\top}
\bm S_{\bm X_{(\ell)}\mid\bm Y_{(\ell)}}
\bm\Phi
\right|.
\label{app:eq:SZ-determinant}
\end{align}
Second, for every positive definite $p\times p$ matrix $\bm S$ and
every orthogonal matrix $(\bm\Phi,\bm\Phi_0)$,
\begin{equation}
\label{app:eq:complement-determinant-identity}
\left|
\bm\Phi_0^{\top}\bm S\bm\Phi_0
\right|
=
|\bm S|
\left|
\bm\Phi^{\top}\bm S^{-1}\bm\Phi
\right|.
\end{equation}
Applying \eqref{app:eq:complement-determinant-identity} to
$\bm S=\widehat{\bm\Sigma}_{\bm X}$ gives
\begin{equation}
\label{app:eq:immaterial-determinant}
\left|
\bm\Phi_0^{\top}
\widehat{\bm\Sigma}_{\bm X}
\bm\Phi_0
\right|
=
\left|
\widehat{\bm\Sigma}_{\bm X}
\right|
\left|
\bm\Phi^{\top}
\widehat{\bm\Sigma}_{\bm X}^{-1}
\bm\Phi
\right|.
\end{equation}
Substituting \eqref{app:eq:SZ-determinant} and
\eqref{app:eq:immaterial-determinant} into
\eqref{app:eq:first-profile-objective}, and discarding all terms that
do not depend on $\bm\Phi$, yields
\begin{equation}
\label{app:eq:profile-objective}
F_q(\bm\Phi)
=
\sum_{\ell=1}^{L}f_\ell
\log
\left|
\bm\Phi^{\top}
\bm S_{\bm X_{(\ell)}\mid\bm Y_{(\ell)}}
\bm\Phi
\right|
+
\log
\left|
\bm\Phi^{\top}
\widehat{\bm\Sigma}_{\bm X}^{-1}
\bm\Phi
\right|.
\end{equation}
This is precisely the objective function in
\eqref{eq:groupwise-predictor-envelope-objective}.

Let $\widehat{\bm\Phi}$ be an orthonormal basis of the minimizing
subspace and let $\widehat{\bm\Phi}_0$ be its orthonormal complement.
Evaluating the profiled estimators above at
$\widehat{\bm\Phi}$ gives the estimators in
\eqref{eq:groupwise-parameter-estimators}--\eqref{eq:groupwise-parameter-estimators_end}.  In particular,
\begin{align}
\widehat{\bm\beta}_{(\ell)}
&=
\widehat{\bm\eta}_{(\ell)}^{\top}
\widehat{\bm\Phi}^{\top}
\nonumber\\
&=
\bm S_{\bm Y_{(\ell)}\bm X_{(\ell)}}
\widehat{\bm\Phi}
\left(
\widehat{\bm\Phi}^{\top}
\bm S_{\bm X_{(\ell)}}
\widehat{\bm\Phi}
\right)^{-1}
\widehat{\bm\Phi}^{\top}.
\label{app:eq:beta-hat}
\end{align}
The groupwise means are estimated by
$\widehat{\bm\mu}_{\bm X_{(\ell)}}=\overline{\bm X}_{(\ell)}$ and
$\widehat{\bm\mu}_{(\ell)}=\overline{\bm Y}_{(\ell)}$, as used when
forming the centered sample covariance matrices at the beginning of
the derivation.

\section{Proof of Theorem~\ref{thm:genv-asymptotic-distribution}}
\label{app:proof-genv-chapter3}

For a positive integer $d$, let $s_d=d(d+1)/2$, and let
$\bm C_d\in\mathbb R^{s_d\times d^2}$ and
$\bm E_d\in\mathbb R^{d^2\times s_d}$ denote the contraction and
expansion matrices, respectively, satisfying
\[
\bm C_d\operatorname{vec}(\bm A)
=
\operatorname{vech}(\bm A),
\qquad
\bm E_d\operatorname{vech}(\bm A)
=
\operatorname{vec}(\bm A),
\qquad
\bm A=\bm A^\top\in\mathbb R^{d\times d}.
\]
For positive integers $a$ and $b$, let
$\bm K_{a,b}\in\mathbb R^{ab\times ab}$ denote the commutation matrix
satisfying
\[
\bm K_{a,b}\operatorname{vec}(\bm A)
=\operatorname{vec}(\bm A^\top),
\qquad
\bm A\in\mathbb R^{a\times b}.
\]
For group $\ell$, consider the identifiable parameter consisting of
the regression coefficient and the predictor covariance matrix.  Stack
these parameters as
\begin{equation}
\label{app:ch3:h}
\bm h(\bm\varphi)
=
\left[
\begin{array}{c}
\operatorname{vec}(\bm\beta_{(1)})\\
\vdots\\
\operatorname{vec}(\bm\beta_{(L)})\\[1mm]
\operatorname{vech}(\bm\Sigma_{\bm X_{(1)}})\\
\vdots\\
\operatorname{vech}(\bm\Sigma_{\bm X_{(L)}})
\end{array}
\right],
\end{equation}
where
\begin{equation}
\label{app:ch3:model-map}
\bm\beta_{(\ell)}
=\bm\eta_{(\ell)}^\top\bm\Phi^\top,
\qquad
\bm\Sigma_{\bm X_{(\ell)}}
=\bm\Phi\bm\Delta_{(\ell)}\bm\Phi^\top
+\bm\Phi_0\bm\Delta_0\bm\Phi_0^\top.
\end{equation}
Here $\bm\eta_{(\ell)}\in\mathbb R^{q\times r}$,
$\bm\Phi\in\mathbb R^{p\times q}$, and
$(\bm\Phi,\bm\Phi_0)$ is orthogonal.  The overparameterized coordinate
vector is
\begin{equation}
\label{app:ch3:varphi}
\bm\varphi
=
\left[
\begin{array}{c}
\operatorname{vec}(\bm\eta_{(1)})\\
\vdots\\
\operatorname{vec}(\bm\eta_{(L)})\\
\operatorname{vec}(\bm\Phi)\\
\operatorname{vech}(\bm\Delta_{(1)})\\
\vdots\\
\operatorname{vech}(\bm\Delta_{(L)})\\
\operatorname{vech}(\bm\Delta_0)
\end{array}
\right].
\end{equation}
The mean parameters and the error covariance matrices
$\bm\Sigma_{(1)},\ldots,\bm\Sigma_{(L)}$ are omitted.  Under joint
normality their score blocks are orthogonal to the blocks in
\eqref{app:ch3:h}; including them therefore does not change the
asymptotic covariance block for the regression coefficients.

For the parameters ordered as in \eqref{app:ch3:h}, the Fisher
information per total observation under the unconstrained model is
\begin{equation}
\label{app:ch3:J}
\bm J
=
\operatorname{blockdiag}
\left\{
\bm J_\beta,\bm J_X
\right\},
\end{equation}
where
\begin{align}
\bm J_\beta
&=
\operatorname{blockdiag}_{\ell=1}^L
\left\{
f_\ell
\left(
\bm\Sigma_{\bm X_{(\ell)}}
\otimes
\bm\Sigma_{(\ell)}^{-1}
\right)
\right\},
\label{app:ch3:J-beta}
\\
\bm J_X
&=
\operatorname{blockdiag}_{\ell=1}^L
\left\{
\frac{f_\ell}{2}
\bm E_p^\top
\left(
\bm\Sigma_{\bm X_{(\ell)}}^{-1}
\otimes
\bm\Sigma_{\bm X_{(\ell)}}^{-1}
\right)
\bm E_p
\right\}.
\label{app:ch3:J-X}
\end{align}

Let
\[
\bm H
=
\frac{\partial\bm h(\bm\varphi)}
{\partial\bm\varphi^\top}
\]
be the structural gradient.  Define
\begin{align*}
\bm A_\eta
&=(\bm\Phi\otimes\bm I_r)\bm K_{q,r},
&
\bm B_{(\ell)}
&=(\bm I_p\otimes\bm\eta_{(\ell)}^\top)\bm K_{p,q},
\\
\bm D_{(\ell)}
&=2\bm C_p
\left(
\bm\Phi\bm\Delta_{(\ell)}\otimes\bm I_p
-\bm\Phi\otimes
\bm\Phi_0\bm\Delta_0\bm\Phi_0^\top
\right),
\\
\bm A_\Delta
&=\bm C_p(\bm\Phi\otimes\bm\Phi)\bm E_q,
&
\bm A_0
&=\bm C_p(\bm\Phi_0\otimes\bm\Phi_0)\bm E_{p-q}.
\end{align*}
Let $\bm 1_L=(1,\ldots,1)^\top\in\mathbb R^L$ denote the vector of
ones.  For compact notation, set
\begin{align*}
\bm{\mathcal A}_\eta
&=\bm I_L\otimes\bm A_\eta,
&
\bm{\mathcal A}_\Delta
&=\bm I_L\otimes\bm A_\Delta,
&
\bm{\mathcal A}_0
&=\bm 1_L\otimes\bm A_0,
\\
\bm{\mathcal B}
&=
\begin{pmatrix}
\bm B_{(1)}\\ \vdots\\ \bm B_{(L)}
\end{pmatrix},
&
\bm{\mathcal D}
&=
\begin{pmatrix}
\bm D_{(1)}\\ \vdots\\ \bm D_{(L)}
\end{pmatrix}.
\end{align*}
The row blocks below correspond to
$\{\operatorname{vec}(\bm\beta_{(\ell)})\}_{\ell=1}^L$ and
$\{\operatorname{vech}(\bm\Sigma_{\bm X_{(\ell)}})\}_{\ell=1}^L$,
whereas the column blocks correspond to
$\{\operatorname{vec}(\bm\eta_{(\ell)})\}_{\ell=1}^L$,
$\operatorname{vec}(\bm\Phi)$,
$\{\operatorname{vech}(\bm\Delta_{(\ell)})\}_{\ell=1}^L$, and
$\operatorname{vech}(\bm\Delta_0)$, respectively.  Hence
\begin{align}
\bm H
&=
\frac{\partial\bm h(\bm\varphi)}
{\partial\bm\varphi^\top}
\nonumber\\
&=
\left[
\begin{array}{c|c|c|c}
\bm{\mathcal A}_\eta
&
\bm{\mathcal B}
&
\bm0
&
\bm0
\\ \hline
\bm0
&
\bm{\mathcal D}
&
\bm{\mathcal A}_\Delta
&
\bm{\mathcal A}_0
\end{array}
\right].
\label{app:ch3:H-explicit}
\end{align}

By Proposition~4.1 of \citet{Shapiro1986},
\begin{equation}
\label{app:ch3:Shapiro-H}
\sqrt n
\left\{
\bm h(\widehat{\bm\varphi})-\bm h(\bm\varphi)
\right\}
\xrightarrow{d}
N(\bm0,\bm V_h),
\qquad
\bm V_h
=
\bm H
(\bm H^\top\bm J\bm H)^\dagger
\bm H^\top.
\end{equation}

As in the single-group proof, replace $\bm H$ by a matrix
$\bm H_1$ with the same column space, chosen so that
$\bm H_1^\top\bm J\bm H_1$ is block diagonal.
Define
\begin{align*}
\widetilde{\bm B}_{(\ell)}
&=
(\bm\Phi_0\otimes\bm\eta_{(\ell)}^\top)
\bm K_{p-q,q},
\\
\widetilde{\bm D}_{(\ell)}
&=
2\bm C_p
\left(
\bm\Phi\bm\Delta_{(\ell)}\otimes\bm\Phi_0
-\bm\Phi\otimes\bm\Phi_0\bm\Delta_0
\right).
\end{align*}
Stack the group-specific blocks as
\begin{align*}
\widetilde{\bm{\mathcal B}}
&=
\begin{pmatrix}
\widetilde{\bm B}_{(1)}\\ \vdots\\
\widetilde{\bm B}_{(L)}
\end{pmatrix},
&
\widetilde{\bm{\mathcal D}}
&=
\begin{pmatrix}
\widetilde{\bm D}_{(1)}\\ \vdots\\
\widetilde{\bm D}_{(L)}
\end{pmatrix}.
\end{align*}
Using the same row ordering as for $\bm H$, and the reduced column
ordering stated above, we obtain
\begin{align}
\bm H_1
&=
\left[
\begin{array}{c|c|c|c}
\bm{\mathcal A}_\eta
&
\widetilde{\bm{\mathcal B}}
&
\bm0
&
\bm0
\\ \hline
\bm0
&
\widetilde{\bm{\mathcal D}}
&
\bm{\mathcal A}_\Delta
&
\bm{\mathcal A}_0
\end{array}
\right].
\label{app:ch3:H1-explicit}
\end{align}

We next give the multi-group analogue of the matrix $\bm H_2$ in the
single-group proof.  Put
\begin{align*}
\bm R_{(\ell)}
&=
\bm K_{q,r}^\top
(\bm\Phi^\top\otimes\bm\eta_{(\ell)}^\top)
\bm K_{p,q},
&
\bm R
&=
\begin{pmatrix}
\bm R_{(1)}\\ \vdots\\ \bm R_{(L)}
\end{pmatrix},
\\
\bm S
&=
\bm K_{p-q,q}^\top
(\bm\Phi_0^\top\otimes\bm I_q)
\bm K_{p,q},
\\
\bm Q_{(\ell)}
&=
2\bm C_q
(\bm\Delta_{(\ell)}\otimes\bm\Phi^\top),
&
\bm Q
&=
\begin{pmatrix}
\bm Q_{(1)}\\ \vdots\\ \bm Q_{(L)}
\end{pmatrix}.
\end{align*}
With column blocks ordered as in \eqref{app:ch3:varphi}, define
\begin{align}
\bm H_2
&=
\left[
\begin{array}{c|c|c|c}
\bm I_{Lqr} & \bm R & \bm0 & \bm0\\
\bm0 & \bm S & \bm0 & \bm0\\
\bm0 & \bm Q & \bm I_{Ls_q} & \bm0\\
\bm0 & \bm0 & \bm0 & \bm I_{s_{p-q}}
\end{array}
\right].
\label{app:ch3:H2}
\end{align}
We now verify the construction by direct matrix multiplication.  From
the displayed forms of $\bm H_1$ and $\bm H_2$,
\begin{align}
\bm H_1\bm H_2
&=
\left[
\begin{array}{c|c|c|c}
\bm{\mathcal A}_\eta
&
\widetilde{\bm{\mathcal B}}
&
\bm0
&
\bm0
\\ \hline
\bm0
&
\widetilde{\bm{\mathcal D}}
&
\bm{\mathcal A}_\Delta
&
\bm{\mathcal A}_0
\end{array}
\right]
\left[
\begin{array}{c|c|c|c}
\bm I_{Lqr} & \bm R & \bm0 & \bm0\\
\bm0 & \bm S & \bm0 & \bm0\\
\bm0 & \bm Q & \bm I_{Ls_q} & \bm0\\
\bm0 & \bm0 & \bm0 & \bm I_{s_{p-q}}
\end{array}
\right]
\nonumber\\
&=
\left[
\begin{array}{c|c|c|c}
\bm{\mathcal A}_\eta
&
\bm{\mathcal A}_\eta\bm R
+\widetilde{\bm{\mathcal B}}\bm S
&
\bm0
&
\bm0
\\ \hline
\bm0
&
\widetilde{\bm{\mathcal D}}\bm S
+\bm{\mathcal A}_\Delta\bm Q
&
\bm{\mathcal A}_\Delta
&
\bm{\mathcal A}_0
\end{array}
\right].
\label{app:ch3:H1H2-first-product}
\end{align}
Therefore, comparison with \eqref{app:ch3:H-explicit} shows that it is
enough to verify, for every $\ell=1,\ldots,L$,
\begin{align}
\bm A_\eta\bm R_{(\ell)}
+\widetilde{\bm B}_{(\ell)}\bm S
&=\bm B_{(\ell)},
\label{app:ch3:H-product-beta-matrix}
\\
\widetilde{\bm D}_{(\ell)}\bm S
+\bm A_\Delta\bm Q_{(\ell)}
&=\bm D_{(\ell)}.
\label{app:ch3:H-product-X-matrix}
\end{align}

We first calculate the left-hand side of
\eqref{app:ch3:H-product-beta-matrix}.  Since
$\bm K_{q,r}\bm K_{q,r}^\top=\bm I_{qr}$,
\begin{align}
\bm A_\eta\bm R_{(\ell)}
&=
(\bm\Phi\otimes\bm I_r)\bm K_{q,r}
\bm K_{q,r}^\top
(\bm\Phi^\top\otimes\bm\eta_{(\ell)}^\top)
\bm K_{p,q}
\nonumber\\
&=
(\bm\Phi\otimes\bm I_r)
(\bm\Phi^\top\otimes\bm\eta_{(\ell)}^\top)
\bm K_{p,q}
\nonumber\\
&=
\left(
\bm\Phi\bm\Phi^\top
\otimes
\bm\eta_{(\ell)}^\top
\right)
\bm K_{p,q}.
\label{app:ch3:AetaR-calculation}
\end{align}
Similarly,
\begin{align}
\widetilde{\bm B}_{(\ell)}\bm S
&=
(\bm\Phi_0\otimes\bm\eta_{(\ell)}^\top)
\bm K_{p-q,q}\bm K_{p-q,q}^\top
(\bm\Phi_0^\top\otimes\bm I_q)
\bm K_{p,q}
\nonumber\\
&=
(\bm\Phi_0\otimes\bm\eta_{(\ell)}^\top)
(\bm\Phi_0^\top\otimes\bm I_q)
\bm K_{p,q}
\nonumber\\
&=
\left(
\bm\Phi_0\bm\Phi_0^\top
\otimes
\bm\eta_{(\ell)}^\top
\right)
\bm K_{p,q}.
\label{app:ch3:BtildeS-calculation}
\end{align}
Adding \eqref{app:ch3:AetaR-calculation} and
\eqref{app:ch3:BtildeS-calculation}, and using
$\bm\Phi\bm\Phi^\top+\bm\Phi_0\bm\Phi_0^\top=\bm I_p$, gives
\begin{align}
\bm A_\eta\bm R_{(\ell)}
+\widetilde{\bm B}_{(\ell)}\bm S
&=
\left[
\left(
\bm\Phi\bm\Phi^\top+\bm\Phi_0\bm\Phi_0^\top
\right)
\otimes\bm\eta_{(\ell)}^\top
\right]\bm K_{p,q}
\nonumber\\
&=
(\bm I_p\otimes\bm\eta_{(\ell)}^\top)\bm K_{p,q}
=\bm B_{(\ell)}.
\label{app:ch3:beta-block-direct-check}
\end{align}
This proves \eqref{app:ch3:H-product-beta-matrix}.

We next verify \eqref{app:ch3:H-product-X-matrix}.  We use
\begin{equation}
\label{app:ch3:commutation-contraction-identities}
\bm K_{a,b}^\top=\bm K_{b,a},
\qquad
\bm C_p\bm K_{p,p}=\bm C_p,
\qquad
\bm E_q\bm C_q
=\frac12(\bm I_{q^2}+\bm K_{q,q}).
\end{equation}
Starting from the definitions of
$\widetilde{\bm D}_{(\ell)}$ and $\bm S$, we have
\begin{align}
\widetilde{\bm D}_{(\ell)}\bm S
={}&
2\bm C_p
\left[
\bm\Phi\bm\Delta_{(\ell)}\otimes\bm\Phi_0
-\bm\Phi\otimes\bm\Phi_0\bm\Delta_0
\right]
\bm K_{p-q,q}^\top
(\bm\Phi_0^\top\otimes\bm I_q)
\bm K_{p,q}
\nonumber\\
={}&
2\bm C_p\bm K_{p,p}
\left[
\bm\Phi_0\otimes\bm\Phi\bm\Delta_{(\ell)}
-\bm\Phi_0\bm\Delta_0\otimes\bm\Phi
\right]
(\bm\Phi_0^\top\otimes\bm I_q)
\bm K_{p,q}
\nonumber\\
={}&
2\bm C_p
\left[
\bm\Phi_0\bm\Phi_0^\top
 \otimes\bm\Phi\bm\Delta_{(\ell)}
-\bm\Phi_0\bm\Delta_0\bm\Phi_0^\top
 \otimes\bm\Phi
\right]
\bm K_{p,q}
\nonumber\\
={}&
2\bm C_p
\left[
\bm\Phi\bm\Delta_{(\ell)}
 \otimes\bm\Phi_0\bm\Phi_0^\top
-\bm\Phi
 \otimes\bm\Phi_0\bm\Delta_0\bm\Phi_0^\top
\right].
\label{app:ch3:DtildeS-calculation}
\end{align}
In the second equality we used the commutation identity
$(\bm A\otimes\bm B)\bm K
=\bm K(\bm B\otimes\bm A)$ with conformable dimensions; in the last
equality we used the same identity once more together with
$\bm C_p\bm K_{p,p}=\bm C_p$.

For the other term, the definitions of $\bm A_\Delta$ and
$\bm Q_{(\ell)}$ give
\begin{align}
\bm A_\Delta\bm Q_{(\ell)}
&=
2\bm C_p
(\bm\Phi\otimes\bm\Phi)
\bm E_q\bm C_q
(\bm\Delta_{(\ell)}\otimes\bm\Phi^\top)
\nonumber\\
&=
\bm C_p
(\bm\Phi\otimes\bm\Phi)
(\bm I_{q^2}+\bm K_{q,q})
(\bm\Delta_{(\ell)}\otimes\bm\Phi^\top)
\nonumber\\
&=
2\bm C_p
\left(
\bm\Phi\bm\Delta_{(\ell)}
\otimes
\bm\Phi\bm\Phi^\top
\right).
\label{app:ch3:AdeltaQ-calculation}
\end{align}
Indeed, the two terms generated by
$\bm I_{q^2}+\bm K_{q,q}$ become equal after premultiplication by
$\bm C_p(\bm\Phi\otimes\bm\Phi)$, because
\[
(\bm\Phi\otimes\bm\Phi)\bm K_{q,q}
=\bm K_{p,p}(\bm\Phi\otimes\bm\Phi),
\qquad
\bm C_p\bm K_{p,p}=\bm C_p.
\]
Adding \eqref{app:ch3:DtildeS-calculation} and
\eqref{app:ch3:AdeltaQ-calculation}, we obtain
\begin{align}
&\widetilde{\bm D}_{(\ell)}\bm S
+\bm A_\Delta\bm Q_{(\ell)}
\nonumber\\
&\quad=
2\bm C_p
\Bigl[
\bm\Phi\bm\Delta_{(\ell)}
\otimes
\left(
\bm\Phi_0\bm\Phi_0^\top
+\bm\Phi\bm\Phi^\top
\right)
-\bm\Phi\otimes
 \bm\Phi_0\bm\Delta_0\bm\Phi_0^\top
\Bigr]
\nonumber\\
&\quad=
2\bm C_p
\left[
\bm\Phi\bm\Delta_{(\ell)}\otimes\bm I_p
-\bm\Phi\otimes
 \bm\Phi_0\bm\Delta_0\bm\Phi_0^\top
\right]
=\bm D_{(\ell)}.
\label{app:ch3:covariance-block-direct-check}
\end{align}
This proves \eqref{app:ch3:H-product-X-matrix}.

Substitution of the two verified identities into
\eqref{app:ch3:H1H2-first-product} yields
\begin{align}
\bm H_1\bm H_2
&=
\left[
\begin{array}{c|c|c|c}
\bm{\mathcal A}_\eta
&\bm{\mathcal B}
&\bm0
&\bm0
\\ \hline
\bm0
&\bm{\mathcal D}
&\bm{\mathcal A}_\Delta
&\bm{\mathcal A}_0
\end{array}
\right]
=\bm H.
\label{app:ch3:H-factorization}
\end{align}

It remains to check that this factorization preserves the column
space.  Since the commutation matrices are orthogonal and
$\bm\Phi_0^\top\bm\Phi_0=\bm I_{p-q}$,
\begin{align}
\bm S\bm S^\top
&=
\bm K_{p-q,q}^\top
(\bm\Phi_0^\top\otimes\bm I_q)
\bm K_{p,q}\bm K_{p,q}^\top
(\bm\Phi_0\otimes\bm I_q)
\bm K_{p-q,q}
\nonumber\\
&=
\bm K_{p-q,q}^\top
\left\{
(\bm\Phi_0^\top\bm\Phi_0)\otimes\bm I_q
\right\}
\bm K_{p-q,q}
=\bm I_{q(p-q)}.
\label{app:ch3:S-full-row-rank}
\end{align}
Thus $\bm S$ has full row rank.  Moreover, if
$[\bm a^\top,\bm b^\top,\bm c^\top,\bm d^\top]\bm H_2=\bm0^\top$,
then direct block multiplication gives
\begin{equation}
\left[
\bm a^\top,
\bm a^\top\bm R+\bm b^\top\bm S+\bm c^\top\bm Q,
\bm c^\top,
\bm d^\top
\right]=\bm0^\top.
\end{equation}
Hence $\bm a=\bm c=\bm d=\bm0$ and
$\bm b^\top\bm S=\bm0^\top$.  The full row rank of $\bm S$ then
implies $\bm b=\bm0$.  Therefore $\bm H_2$ has full row rank and,
consequently, it has a right inverse $\bm H_2^{\mathrm R}$.  Since
$\bm H=\bm H_1\bm H_2$ and
$\bm H_1=\bm H\bm H_2^{\mathrm R}$, we obtain
\begin{equation}
\label{app:ch3:span}
\operatorname{span}(\bm H)
=\operatorname{span}(\bm H_1).
\end{equation}
Since the covariance in \eqref{app:ch3:Shapiro-H} depends on
$\bm H$ only through its column space,
\begin{equation}
\label{app:ch3:Shapiro-H1}
\bm V_h
=
\bm H_1
(\bm H_1^\top\bm J\bm H_1)^\dagger
\bm H_1^\top.
\end{equation}

It remains to calculate $\bm H_1^\top\bm J\bm H_1$.  Using the
displayed block forms of $\bm H_1$ and
$\bm J=\operatorname{blockdiag}(\bm J_\beta,\bm J_X)$, direct
multiplication gives
\begin{align}
\bm H_1^\top\bm J\bm H_1
&=
\left[
\begin{array}{c|c|c|c}
\bm{\mathcal A}_\eta^\top\bm J_\beta\bm{\mathcal A}_\eta
&
\bm{\mathcal A}_\eta^\top\bm J_\beta
 \widetilde{\bm{\mathcal B}}
&
\bm0
&
\bm0
\\ \hline
\widetilde{\bm{\mathcal B}}^\top\bm J_\beta
 \bm{\mathcal A}_\eta
&
\widetilde{\bm{\mathcal B}}^\top\bm J_\beta
 \widetilde{\bm{\mathcal B}}
+\widetilde{\bm{\mathcal D}}^\top\bm J_X
 \widetilde{\bm{\mathcal D}}
&
\widetilde{\bm{\mathcal D}}^\top\bm J_X
 \bm{\mathcal A}_\Delta
&
\widetilde{\bm{\mathcal D}}^\top\bm J_X
 \bm{\mathcal A}_0
\\ \hline
\bm0
&
\bm{\mathcal A}_\Delta^\top\bm J_X
 \widetilde{\bm{\mathcal D}}
&
\bm{\mathcal A}_\Delta^\top\bm J_X
 \bm{\mathcal A}_\Delta
&
\bm{\mathcal A}_\Delta^\top\bm J_X
 \bm{\mathcal A}_0
\\ \hline
\bm0
&
\bm{\mathcal A}_0^\top\bm J_X
 \widetilde{\bm{\mathcal D}}
&
\bm{\mathcal A}_0^\top\bm J_X
 \bm{\mathcal A}_\Delta
&
\bm{\mathcal A}_0^\top\bm J_X
 \bm{\mathcal A}_0
\end{array}
\right].
\label{app:ch3:H1JH1-direct}
\end{align}

The off-diagonal blocks in
\eqref{app:ch3:H1JH1-direct} vanish.  For each $\ell$,
\begin{align}
\bm A_\eta^\top\bm J_{\beta,\ell}
\widetilde{\bm B}_{(\ell)}
&=
f_\ell\bm K_{q,r}^\top
\left[
\bm\Phi^\top\bm\Sigma_{\bm X_{(\ell)}}\bm\Phi_0
\otimes
\bm\Sigma_{(\ell)}^{-1}\bm\eta_{(\ell)}^\top
\right]\bm K_{p-q,q}
=\bm0,
\label{app:ch3:eta-Phi-cross-zero}
\end{align}
because
$\bm\Phi^\top\bm\Sigma_{\bm X_{(\ell)}}\bm\Phi_0=\bm0$.
The other off-diagonal blocks satisfy
\begin{align}
\widetilde{\bm D}_{(\ell)}^\top
\bm J_{X,\ell}\bm A_\Delta
&=\bm0,
&
\widetilde{\bm D}_{(\ell)}^\top
\bm J_{X,\ell}\bm A_0
&=\bm0,
&
\bm A_\Delta^\top\bm J_{X,\ell}\bm A_0
&=\bm0.
\label{app:ch3:covariance-cross-zero}
\end{align}
Indeed, after inserting
\[
\bm\Sigma_{\bm X_{(\ell)}}^{-1}
=
\bm\Phi\bm\Delta_{(\ell)}^{-1}\bm\Phi^\top
+\bm\Phi_0\bm\Delta_0^{-1}\bm\Phi_0^\top,
\]
each product in \eqref{app:ch3:covariance-cross-zero} contains either
$\bm\Phi^\top\bm\Phi_0$ or $\bm\Phi_0^\top\bm\Phi$, and is therefore
zero.

We now evaluate the two diagonal blocks that enter the asymptotic
variance of $\widehat{\bm\beta}_{(\ell)}$.  First,
\begin{align}
\bm I_{\eta,\ell}
&:=
\bm A_\eta^\top\bm J_{\beta,\ell}\bm A_\eta
\nonumber\\
&=
f_\ell\bm K_{q,r}^\top
(\bm\Phi^\top\otimes\bm I_r)
\left(
\bm\Sigma_{\bm X_{(\ell)}}
\otimes\bm\Sigma_{(\ell)}^{-1}
\right)
(\bm\Phi\otimes\bm I_r)
\bm K_{q,r}
\nonumber\\
&=
f_\ell\bm K_{q,r}^\top
\left[
\left(
\bm\Phi^\top\bm\Sigma_{\bm X_{(\ell)}}\bm\Phi
\right)
\otimes\bm\Sigma_{(\ell)}^{-1}
\right]
\bm K_{q,r}
\nonumber\\
&=
f_\ell\bm K_{q,r}^\top
\left(
\bm\Delta_{(\ell)}
\otimes\bm\Sigma_{(\ell)}^{-1}
\right)
\bm K_{q,r}.
\label{app:ch3:eta-information}
\end{align}

Next, the $(2,2)$ block of \eqref{app:ch3:H1JH1-direct} is
\begin{align}
\bm I_\Phi
&:=
\widetilde{\bm{\mathcal B}}^\top\bm J_\beta
\widetilde{\bm{\mathcal B}}
+
\widetilde{\bm{\mathcal D}}^\top\bm J_X
\widetilde{\bm{\mathcal D}}
\nonumber\\
&=
\sum_{k=1}^L
\left\{
\widetilde{\bm B}_{(k)}^\top
\bm J_{\beta,k}
\widetilde{\bm B}_{(k)}
+
\widetilde{\bm D}_{(k)}^\top
\bm J_{X,k}
\widetilde{\bm D}_{(k)}
\right\}.
\label{app:ch3:Phi-information-direct}
\end{align}
The first product in the braces is
\begin{align}
\widetilde{\bm B}_{(\ell)}^\top
\bm J_{\beta,\ell}
\widetilde{\bm B}_{(\ell)}
&=
f_\ell\bm K_{p-q,q}^\top
\left(
\bm\Phi_0^\top\otimes\bm\eta_{(\ell)}
\right)
\left(
\bm\Sigma_{\bm X_{(\ell)}}
\otimes
\bm\Sigma_{(\ell)}^{-1}
\right)
\left(
\bm\Phi_0\otimes\bm\eta_{(\ell)}^\top
\right)
\bm K_{p-q,q}
\nonumber\\
&=
f_\ell\bm K_{p-q,q}^\top
\left[
\left(
\bm\Phi_0^\top
\bm\Sigma_{\bm X_{(\ell)}}
\bm\Phi_0
\right)
\otimes
\bm\eta_{(\ell)}
\bm\Sigma_{(\ell)}^{-1}
\bm\eta_{(\ell)}^\top
\right]
\bm K_{p-q,q}
\nonumber\\
&=
f_\ell\bm K_{p-q,q}^\top
\left[
\left\{
\bm\Phi_0^\top
\left(
\bm\Phi\bm\Delta_{(\ell)}\bm\Phi^\top
+
\bm\Phi_0\bm\Delta_0\bm\Phi_0^\top
\right)
\bm\Phi_0
\right\}
\otimes
\bm\eta_{(\ell)}
\bm\Sigma_{(\ell)}^{-1}
\bm\eta_{(\ell)}^\top
\right]
\bm K_{p-q,q}
\nonumber\\
&=
f_\ell\bm K_{p-q,q}^\top
\left[
\bm\Delta_0
\otimes
\bm\eta_{(\ell)}
\bm\Sigma_{(\ell)}^{-1}
\bm\eta_{(\ell)}^\top
\right]\bm K_{p-q,q}.
\label{app:ch3:Phi-information-beta-product}
\end{align}
For the second product, write
\[
\bm G_{(\ell)}
=
\left(
\bm\Phi\bm\Delta_{(\ell)}\otimes\bm\Phi_0
-\bm\Phi\otimes\bm\Phi_0\bm\Delta_0
\right),
\qquad
\widetilde{\bm D}_{(\ell)}=2\bm C_p\bm G_{(\ell)}.
\]
Substituting $\widetilde{\bm D}_{(\ell)}$ and $\bm J_{X,\ell}$ and
using $\bm E_p\bm C_p=(\bm I_{p^2}+\bm K_{p,p})/2$ gives
\begin{align}
\widetilde{\bm D}_{(\ell)}^\top
\bm J_{X,\ell}
\widetilde{\bm D}_{(\ell)}
&=
\left(2\bm C_p\bm G_{(\ell)}\right)^\top
\left[
\frac{f_\ell}{2}\bm E_p^\top
\left(
\bm\Sigma_{\bm X_{(\ell)}}^{-1}
\otimes
\bm\Sigma_{\bm X_{(\ell)}}^{-1}
\right)
\bm E_p
\right]
\left(2\bm C_p\bm G_{(\ell)}\right)
\nonumber\\
&=
2f_\ell\bm G_{(\ell)}^\top
\left(\bm E_p\bm C_p\right)^\top
\left(
\bm\Sigma_{\bm X_{(\ell)}}^{-1}
\otimes
\bm\Sigma_{\bm X_{(\ell)}}^{-1}
\right)
\left(\bm E_p\bm C_p\right)\bm G_{(\ell)}
\nonumber\\
&=
\frac{f_\ell}{2}\bm G_{(\ell)}^\top
\left(\bm I_{p^2}+\bm K_{p,p}\right)
\left(
\bm\Sigma_{\bm X_{(\ell)}}^{-1}
\otimes
\bm\Sigma_{\bm X_{(\ell)}}^{-1}
\right)
\left(\bm I_{p^2}+\bm K_{p,p}\right)
\bm G_{(\ell)}
\nonumber\\
&=
f_\ell\bm G_{(\ell)}^\top
\left(
\bm\Sigma_{\bm X_{(\ell)}}^{-1}
\otimes
\bm\Sigma_{\bm X_{(\ell)}}^{-1}
\right)
\left(\bm I_{p^2}+\bm K_{p,p}\right)
\bm G_{(\ell)}
\nonumber\\
&=
f_\ell\bm G_{(\ell)}^\top
\left(
\bm\Sigma_{\bm X_{(\ell)}}^{-1}
\otimes
\bm\Sigma_{\bm X_{(\ell)}}^{-1}
\right)
\bm G_{(\ell)}
\nonumber\\
&=
f_\ell
\Bigl[
\bm\Delta_{(\ell)}
\bm\Phi^\top
\bm\Sigma_{\bm X_{(\ell)}}^{-1}
\bm\Phi
\bm\Delta_{(\ell)}
\otimes
\bm\Phi_0^\top
\bm\Sigma_{\bm X_{(\ell)}}^{-1}
\bm\Phi_0
\nonumber\\
&\qquad-
\bm\Delta_{(\ell)}
\bm\Phi^\top
\bm\Sigma_{\bm X_{(\ell)}}^{-1}
\bm\Phi
\otimes
\bm\Phi_0^\top
\bm\Sigma_{\bm X_{(\ell)}}^{-1}
\bm\Phi_0\bm\Delta_0
\nonumber\\
&\qquad-
\bm\Phi^\top
\bm\Sigma_{\bm X_{(\ell)}}^{-1}
\bm\Phi\bm\Delta_{(\ell)}
\otimes
\bm\Delta_0\bm\Phi_0^\top
\bm\Sigma_{\bm X_{(\ell)}}^{-1}
\bm\Phi_0
\nonumber\\
&\qquad+
\bm\Phi^\top
\bm\Sigma_{\bm X_{(\ell)}}^{-1}
\bm\Phi
\otimes
\bm\Delta_0\bm\Phi_0^\top
\bm\Sigma_{\bm X_{(\ell)}}^{-1}
\bm\Phi_0\bm\Delta_0
\Bigr]
\nonumber\\
&=
f_\ell
\left[
\bm\Delta_{(\ell)}\otimes\bm\Delta_0^{-1}
+
\bm\Delta_{(\ell)}^{-1}\otimes\bm\Delta_0
-
2\bm I_q\otimes\bm I_{p-q}
\right]
\nonumber\\
&=
f_\ell\bm K_{p-q,q}^\top
\Bigl[
\bm\Delta_0^{-1}\otimes\bm\Delta_{(\ell)}
+\bm\Delta_0\otimes\bm\Delta_{(\ell)}^{-1}
-2\bm I_{p-q}\otimes\bm I_q
\Bigr]\bm K_{p-q,q}.
\label{app:ch3:Phi-information-X-product}
\end{align}
The reduction from two factors
$\bm I_{p^2}+\bm K_{p,p}$ to one uses
$\bm K_{p,p}^2=\bm I_{p^2}$ and the fact that
$\bm K_{p,p}$ commutes with
$\bm\Sigma_{\bm X_{(\ell)}}^{-1}\otimes
\bm\Sigma_{\bm X_{(\ell)}}^{-1}$.
The remaining term containing $\bm K_{p,p}$ then vanishes because
each of its Kronecker products contains either
$\bm\Phi^\top\bm\Sigma_{\bm X_{(\ell)}}^{-1}\bm\Phi_0$ or
$\bm\Phi_0^\top\bm\Sigma_{\bm X_{(\ell)}}^{-1}\bm\Phi$.
The final equality uses the commutation identity to reverse the order
of the two Kronecker factors.

To combine \eqref{app:ch3:Phi-information-beta-product} and
\eqref{app:ch3:Phi-information-X-product}, define
\begin{align}
\bm T_{(\ell)}
={}&
\bm\Delta_0
\otimes
\bm\eta_{(\ell)}
\bm\Sigma_{(\ell)}^{-1}
\bm\eta_{(\ell)}^\top
\nonumber\\
&+
\bm\Delta_0^{-1}\otimes\bm\Delta_{(\ell)}
+\bm\Delta_0\otimes\bm\Delta_{(\ell)}^{-1}
-2\bm I_{p-q}\otimes\bm I_q,
\label{app:ch3:T-ell}
\\
\bm M
&=
\sum_{k=1}^L f_k\bm T_{(k)}.
\label{app:ch3:M}
\end{align}
Then \eqref{app:ch3:Phi-information-direct} becomes
\begin{equation}
\label{app:ch3:Phi-information}
\bm I_\Phi
=
\bm K_{p-q,q}^\top\bm M\bm K_{p-q,q}.
\end{equation}

Consequently,
\begin{equation}
\label{app:ch3:block-information}
\bm H_1^\top\bm J\bm H_1
=
\operatorname{blockdiag}
\left\{
\{\bm I_{\eta,\ell}\}_{\ell=1}^L,
\bm K_{p-q,q}^\top\bm M\bm K_{p-q,q},
\bm I_{\mathrm{nuis}}
\right\},
\end{equation}
where $\bm I_{\mathrm{nuis}}$ collects the mutually orthogonal
$\bm\Delta_{(1)},\ldots,\bm\Delta_{(L)},\bm\Delta_0$ blocks, which do
not enter the $\bm\beta_{(\ell)}$ variance block.

Let $\bm H_{1,\beta_\ell}$ denote the rows of $\bm H_1$ corresponding
to $\operatorname{vec}(\bm\beta_{(\ell)})$.  From
\eqref{app:ch3:H1-explicit},
\begin{equation}
\label{app:ch3:H1-beta-row}
\bm H_{1,\beta_\ell}
=
\left[
\begin{array}{c|c|c|c}
\bm0\ \cdots\ \bm A_\eta\ \cdots\ \bm0
&
\widetilde{\bm B}_{(\ell)}
&
\bm0
&
\bm0
\end{array}
\right].
\end{equation}
Hence the $rp\times rp$ diagonal block of $\bm H_1(\bm H_1^\top\bm J\bm H_1)^\dagger\bm H_1^\top$ corresponding to $\operatorname{vec}(\bm\beta_{(\ell)})$ is
\begin{align}
&\bm H_{1,\beta_\ell}
(\bm H_1^\top\bm J\bm H_1)^\dagger
\bm H_{1,\beta_\ell}^\top
\nonumber\\
&\qquad=
\bm A_\eta\bm I_{\eta,\ell}^{-1}\bm A_\eta^\top
+
\widetilde{\bm B}_{(\ell)}
\bm I_\Phi^\dagger
\widetilde{\bm B}_{(\ell)}^\top.
\label{app:ch3:beta-block-extraction}
\end{align}
Substituting the definitions of $\bm A_\eta$,
$\widetilde{\bm B}_{(\ell)}$, and $\bm I_\Phi$ gives
\begin{align}
\operatorname{avar}
\left\{
\sqrt n\operatorname{vec}
(\widehat{\bm\beta}_{(\ell),\mathrm{genv}})
\right\}
={}&
(\bm\Phi\otimes\bm I_r)\bm K_{q,r}
\bm I_{\eta,\ell}^{-1}
\bm K_{q,r}^\top
(\bm\Phi^\top\otimes\bm I_r)
\nonumber\\
&+
(\bm\Phi_0\otimes\bm\eta_{(\ell)}^\top)\bm K_{p-q,q}
(\bm K_{p-q,q}^\top\bm M\bm K_{p-q,q})^\dagger
\bm K_{p-q,q}^\top
(\bm\Phi_0^\top\otimes\bm\eta_{(\ell)}).
\label{app:ch3:beta-block-before}
\end{align}
Since commutation matrices are orthogonal permutation matrices,
\[
(\bm K_{p-q,q}^\top\bm M\bm K_{p-q,q})^\dagger
=\bm K_{p-q,q}^\top\bm M^\dagger\bm K_{p-q,q}.
\]
Using this identity, \eqref{app:ch3:eta-information}, and the
mixed-product rule for Kronecker products, equation
\eqref{app:ch3:beta-block-before} becomes
\begin{align}
\operatorname{avar}
\left\{
\sqrt n\operatorname{vec}
(\widehat{\bm\beta}_{(\ell),\mathrm{genv}})
\right\}
={}&
f_\ell^{-1}
\left(
\bm\Phi\bm\Delta_{(\ell)}^{-1}\bm\Phi^\top
\otimes
\bm\Sigma_{(\ell)}
\right)
\nonumber\\
&+
(\bm\Phi_0\otimes\bm\eta_{(\ell)}^\top)
\bm M^\dagger
(\bm\Phi_0^\top\otimes\bm\eta_{(\ell)}).
\label{app:ch3:beta-avar}
\end{align}
The right-hand side of \eqref{app:ch3:beta-avar} is
$\bm V_{(\ell),\mathrm{genv}}$.  Hence
\[
\sqrt n
\left\{
\operatorname{vec}
(\widehat{\bm\beta}_{(\ell),\mathrm{genv}})
-\operatorname{vec}(\bm\beta_{(\ell)})
\right\}
\xrightarrow{d}
N(\bm0,\bm V_{(\ell),\mathrm{genv}}).
\]
It remains to verify the discrepancy-function assumptions used in
\eqref{app:ch3:Shapiro-H}.

\subsection{Verification of Shapiro's conditions for the discrepancy function}
\label{app:shapiro-discrepancy}

We verify Conditions 1--4 in Section~3 of \citet{Shapiro1986} for
the likelihood discrepancy used here.

For group $\ell$, let
$\widehat{\bm\beta}_{(\ell),\mathrm{std}}$,
$\widehat{\bm\Sigma}_{(\ell)}$, and
$\widehat{\bm\Sigma}_{\bm X_{(\ell)}}$ be the Gaussian maximum
likelihood estimators under the standard model.  Let
\[
 \bm x=
 \left\{
 \widehat{\bm\beta}_{(\ell),\mathrm{std}},
 \widehat{\bm\Sigma}_{(\ell)},
 \widehat{\bm\Sigma}_{\bm X_{(\ell)}}
 \right\}_{\ell=1}^L
\]
and let
\[
 \bm\xi=
 \left\{
 \bm\beta_{(\ell)},
 \bm\Sigma_{(\ell)},
 \bm\Sigma_{\bm X_{(\ell)}}
 \right\}_{\ell=1}^L
\]
be a candidate parameter value.  All covariance matrices are assumed
to be positive definite.  The variables are centered, so mean
parameters do not appear below.  The error covariance matrices are
included because they occur in the full Gaussian likelihood.

For positive definite $d\times d$ matrices $\bm S$ and $\bm\Omega$,
define
\begin{equation}
 \label{app:matrix-discrepancy}
 d_d(\bm S,\bm\Omega)
 =
 \log|\bm\Omega|-\log|\bm S|
 +\operatorname{tr}(\bm\Omega^{-1}\bm S)-d.
\end{equation}
Let $l_{\max}$ be the maximum of the standard-model log likelihood,
and let $l(\bm\xi)$ be that likelihood evaluated at $\bm\xi$.
Completing the square in the regression coefficients, the discrepancy
function can be written as
\begin{equation}
 \label{app:likelihood-discrepancy}
\begin{split}
 F_n(\bm x,\bm\xi)
 &:={\frac{2}{n}}\{l_{\max}-l(\bm\xi)\}\\
 &=\sum_{\ell=1}^L f_\ell
 \Bigl[
 d_r\!\left(
 \widehat{\bm\Sigma}_{(\ell)},
 \bm\Sigma_{(\ell)}
 \right)
 +d_p\!\left(
 \widehat{\bm\Sigma}_{\bm X_{(\ell)}},
 \bm\Sigma_{\bm X_{(\ell)}}
 \right)\\
 &\hspace{35mm}
 +\operatorname{tr}\!\left\{
 \bm\Sigma_{(\ell)}^{-1}
 \bm A_{(\ell)}
 \widehat{\bm\Sigma}_{\bm X_{(\ell)}}
 \bm A_{(\ell)}^\top
 \right\}
 \Bigr],
\end{split}
\end{equation}
where
\[
 \bm A_{(\ell)}
 =\bm\beta_{(\ell)}
 -\widehat{\bm\beta}_{(\ell),\mathrm{std}}.
\]
The term involving $d_p$ appears because the likelihood includes the
marginal distribution of $\bm X$.  It is absent from the fixed-design
likelihood in Park et al.\ (2017).  Multiplication by $2/n$ does not
change the minimizer of $l_{\max}-l(\bm\xi)$.  With this scaling, one
half of the Hessian of $F_n$ at the true parameter is the Fisher
information per observation, $\bm J$.

\paragraph{Condition 1.}
Set
$\bm R=\bm\Omega^{-1/2}\bm S\bm\Omega^{-1/2}$, and let
$\lambda_1,\ldots,\lambda_d>0$ be its eigenvalues.  Since
$|\bm R|=|\bm S|/|\bm\Omega|$ and
$\operatorname{tr}(\bm R)=\operatorname{tr}(\bm\Omega^{-1}\bm S)$,
\begin{align}
 d_d(\bm S,\bm\Omega)
 &=-\log|\bm R|+\operatorname{tr}(\bm R)-d
 \nonumber\\
 &=\sum_{j=1}^d
 \{\lambda_j-\log\lambda_j-1\}
 \geq0.
 \label{app:stein-loss-positive}
\end{align}
The inequality follows from
$a-\log a-1\geq0$ for every $a>0$.  In addition,
\begin{equation}
\label{app:regression-loss-positive}
\operatorname{tr}\!\left\{\bm\Sigma_{(\ell)}^{-1}\bm A_{(\ell)}\widehat{\bm\Sigma}_{\bm X_{(\ell)}}\bm A_{(\ell)}^\top\right\}
=\left\|\bm\Sigma_{(\ell)}^{-1/2}\bm A_{(\ell)}\widehat{\bm\Sigma}_{\bm X_{(\ell)}}^{1/2}\right\|_F^2\geq0.
\end{equation}
Thus every term in \eqref{app:likelihood-discrepancy} is nonnegative.
Since $f_\ell>0$, it follows that
$F_n(\bm x,\bm\xi)\geq0$, which proves Condition 1.

\paragraph{Condition 2.}
Since $f_\ell>0$ and every term in
\eqref{app:likelihood-discrepancy} is nonnegative,
$F_n(\bm x,\bm\xi)=0$ if and only if every term in that expression
is zero.

First, equality in \eqref{app:stein-loss-positive} holds if and only
if $\lambda_j=1$ for every $j$.  Since
$\bm\Omega^{-1/2}\bm S\bm\Omega^{-1/2}$ is symmetric, this is
equivalent to
\[
\bm\Omega^{-1/2}\bm S\bm\Omega^{-1/2}=\bm I_d,
\]
and hence to $\bm S=\bm\Omega$.  Therefore,
\[
d_d(\bm S,\bm\Omega)=0
\quad\Longleftrightarrow\quad
\bm S=\bm\Omega.
\]

Next, by \eqref{app:regression-loss-positive}, the term involving
$\bm A_{(\ell)}$ is zero if and only if
\[
\bm\Sigma_{(\ell)}^{-1/2}
\bm A_{(\ell)}
\widehat{\bm\Sigma}_{\bm X_{(\ell)}}^{1/2}
=\bm0.
\]
Both covariance matrices are positive definite, so
$\bm\Sigma_{(\ell)}^{-1/2}$ and
$\widehat{\bm\Sigma}_{\bm X_{(\ell)}}^{1/2}$ are nonsingular.
Consequently, the preceding equality holds if and only if
\[
\bm A_{(\ell)}=\bm0
\quad\Longleftrightarrow\quad
\bm\beta_{(\ell)}
=\widehat{\bm\beta}_{(\ell),\mathrm{std}}.
\]

It follows that $F_n(\bm x,\bm\xi)=0$ if and only if, for every
$\ell$,
\[
 \bm\beta_{(\ell)}
 =\widehat{\bm\beta}_{(\ell),\mathrm{std}},\qquad
 \bm\Sigma_{(\ell)}
 =\widehat{\bm\Sigma}_{(\ell)},\qquad
 \bm\Sigma_{\bm X_{(\ell)}}
 =\widehat{\bm\Sigma}_{\bm X_{(\ell)}}.
\]
These equalities are precisely $\bm x=\bm\xi$.  Conversely, if
$\bm x=\bm\xi$, every term in
\eqref{app:likelihood-discrepancy} is zero.  Thus
\[
F_n(\bm x,\bm\xi)=0
\quad\Longleftrightarrow\quad
\bm x=\bm\xi,
\]
which proves Condition 2.

\paragraph{Condition 3.}
On the set of positive definite matrices,
$\log|\bm\Omega|$ and $\bm\Omega^{-1}$ are infinitely continuously
differentiable functions of $\bm\Omega$.  All other terms in
\eqref{app:likelihood-discrepancy} are polynomial or bilinear.
Therefore $F_n(\bm x,\bm\xi)$ is twice continuously differentiable
in $\bm x$ and $\bm\xi$.

\paragraph{Condition 4.}
This condition requires constants $M>0$ and $\varepsilon>0$ such that
\begin{equation}
 \label{app:shapiro-condition-four}
 F_n(\bm x,\bm\xi)\geq\varepsilon
 \quad\text{whenever}\quad
 \|\bm x-\bm\xi\|\geq M.
\end{equation}
It does not hold on the entire positive definite cone.  To see this,
set all corresponding components of $\bm x_t$ and $\bm\xi_t$ equal
except for one error covariance matrix, and let
\[
 \widehat{\bm\Sigma}_{(1),t}=t\bm I_r,
 \qquad
 \bm\Sigma_{(1),t}=(t+\sqrt t)\bm I_r.
\]
Then
\[
 \left\|
 \operatorname{vech}(\widehat{\bm\Sigma}_{(1),t})
 -\operatorname{vech}(\bm\Sigma_{(1),t})
 \right\|
 =\sqrt{rt}\longrightarrow\infty,
\]
whereas
\begin{equation}
 \label{app:condition-four-counterexample}
 F_n(\bm x_t,\bm\xi_t)
 =f_1 r
 \left\{
 \log(1+t^{-1/2})
 +(1+t^{-1/2})^{-1}-1
 \right\}
 \longrightarrow0.
\end{equation}
Thus the distance between $\bm x_t$ and $\bm\xi_t$ tends to infinity
while the discrepancy tends to zero.  Therefore Conditions 1--3 hold
on the positive definite cone, but Condition 4 does not.

We therefore restrict $\bm x$ and $\bm\xi$ to a compact set
$\mathcal K$ containing the true parameter in its interior.  Assume
that, for constants $0<c<C<\infty$ and $B<\infty$,
\begin{equation}
 \label{app:compact-localization}
 c\bm I\preceq
 \bm\Sigma_{(\ell)},
 \widehat{\bm\Sigma}_{(\ell)},
 \bm\Sigma_{\bm X_{(\ell)}},
 \widehat{\bm\Sigma}_{\bm X_{(\ell)}}
 \preceq C\bm I,
 \qquad
 \|\bm\beta_{(\ell)}\|_F,
 \|\widehat{\bm\beta}_{(\ell),\mathrm{std}}\|_F
 \leq B.
\end{equation}
Choose $M>0$ such that
\[
 \mathcal A_M
 =\{(\bm x,\bm\xi)\in\mathcal K^2:
       \|\bm x-\bm\xi\|\geq M\}
\]
is nonempty.  The set $\mathcal A_M$ is compact and does not contain
any point with $\bm x=\bm\xi$.  By Conditions 1--3, the unweighted
sum of the bracketed terms in
\eqref{app:likelihood-discrepancy} is continuous and strictly positive
on $\mathcal A_M$.  Hence its minimum on $\mathcal A_M$, denoted by
$\delta_M$, is positive.

Let
$\underline f=\min_{1\leq\ell\leq L}f_\ell>0$.
Consequently,
\[
 F_n(\bm x,\bm\xi)
 \geq \underline f\delta_M
\]
whenever $(\bm x,\bm\xi)\in\mathcal A_M$.  Thus
\eqref{app:shapiro-condition-four} holds on $\mathcal K$ with
$\varepsilon=\underline f\delta_M$.

Conditions 1--4 therefore hold on $\mathcal K$.  Since the true
parameter is an interior point of $\mathcal K$ and the standard-model
estimator is consistent, this restriction does not change the
first-order asymptotic distribution.  Finally, because $\bm J$ is
positive definite,
\[
\operatorname{rank}(\bm H^\top\bm J\bm H)
=\operatorname{rank}(\bm H),
\]
and the regular-point assumption ensures that the rank of $\bm H$ is
constant in a neighborhood of the true parameter.  Proposition~4.1
of \citet{Shapiro1986} is therefore applicable, completing the proof.
\hfill$\square$

\section{Proof of Corollary~\ref{cor:asymptotic-efficiency}}
\label{app:proof-efficiency}

In this section, we compare the asymptotic covariance matrices in
\eqref{eq:avar-genv} and \eqref{eq:avar-sep}.  Their first terms are
identical, so it suffices to compare the terms involving $\bm M$ and
$\bm M_{(\ell)}$.  We first show that
$\bm M\succeq\bm M_{(\ell)}$ and then use this ordering to compare
the corresponding quadratic forms involving their Moore--Penrose
inverses.

Adding \eqref{app:ch3:Phi-information-beta-product} and
\eqref{app:ch3:Phi-information-X-product} gives
\[
f_k\bm K_{p-q,q}^\top\bm T_{(k)}\bm K_{p-q,q}
=
\widetilde{\bm B}_{(k)}^\top
\bm J_{\beta,k}\widetilde{\bm B}_{(k)}
+
\widetilde{\bm D}_{(k)}^\top
\bm J_{X,k}\widetilde{\bm D}_{(k)}
\succeq\bm0.
\]
Since $f_k>0$ and $\bm K_{p-q,q}$ is nonsingular,
$\bm T_{(k)}\succeq\bm0$.  By \eqref{eq:T-matrix} and
\eqref{app:ch3:T-ell},
\[
\bm M_{(\ell)}=f_\ell\bm T_{(\ell)}.
\]
Therefore, from \eqref{app:ch3:M},
\begin{equation}
\label{app:eq:M-order}
\bm M
=
\bm M_{(\ell)}
+
\sum_{k\ne\ell}f_k\bm T_{(k)}
\succeq
\bm M_{(\ell)}.
\end{equation}

To use \eqref{app:eq:M-order} when the matrices may be singular, we
first establish an inequality for quadratic forms involving their
Moore--Penrose inverses.  For a matrix
$\bm C\in\mathbb R^{m\times d}$, define its column space and null
space by
\[
\mathcal R(\bm C)
=
\{\bm C\bm z:\bm z\in\mathbb R^d\},
\qquad
\mathcal N(\bm C)
=
\{\bm z\in\mathbb R^d:\bm C\bm z=\bm0\}.
\]

Suppose that $\bm0\preceq\bm A\preceq\bm B$ and
$\bm y\in\mathcal R(\bm A)$.  The ordering implies
$\mathcal N(\bm B)\subseteq\mathcal N(\bm A)$ and hence
$\mathcal R(\bm A)\subseteq\mathcal R(\bm B)$, so
$\bm y\in\mathcal R(\bm C)$ for $\bm C=\bm A$ and $\bm B$.
Since $\bm C\bm C^\dagger\bm y=\bm y$, completing the square gives
\[
2\bm y^\top\bm z-\bm z^\top\bm C\bm z=\bm y^\top\bm C^\dagger\bm y-(\bm z-\bm C^\dagger\bm y)^\top\bm C(\bm z-\bm C^\dagger\bm y).
\]
The final term is nonpositive and equals zero at
$\bm z=\bm C^\dagger\bm y$; therefore,
\[
\max_{\bm z}\{2\bm y^\top\bm z-\bm z^\top\bm C\bm z\}=\bm y^\top\bm C^\dagger\bm y.
\]
Since $\bm A\preceq\bm B$, comparison of the two maxima gives
\[
\bm y^\top\bm B^\dagger\bm y=\max_{\bm z}\{2\bm y^\top\bm z-\bm z^\top\bm B\bm z\}\leq\max_{\bm z}\{2\bm y^\top\bm z-\bm z^\top\bm A\bm z\}=\bm y^\top\bm A^\dagger\bm y.
\]

To apply this result with $\bm A=\bm M_{(\ell)}$ and
$\bm B=\bm M$, it remains to verify that the vectors appearing in
the relevant quadratic forms belong to
$\mathcal R(\bm M_{(\ell)})$.  For subspaces
$\mathcal U\subseteq\mathbb R^m$ and
$\mathcal V\subseteq\mathbb R^p$, define
\[
\mathcal U\otimes\mathcal V
=
\operatorname{span}
\{\bm u\otimes\bm v:
\bm u\in\mathcal U,\ \bm v\in\mathcal V\}.
\]
Since
$\bm\Phi_0^\top$ has full row rank and $\bm\Delta_0$ is positive
definite,
\[
\mathcal R(\bm\Phi_0^\top)
=
\mathbb R^{p-q}
=
\mathcal R(\bm\Delta_0).
\]
Moreover, setting
$\bm G=\bm\eta_{(\ell)}\bm\Sigma_{(\ell)}^{-1/2}$ gives
\[
\begin{aligned}
\mathcal R\!\left(
\bm\eta_{(\ell)}
\bm\Sigma_{(\ell)}^{-1}
\bm\eta_{(\ell)}^\top
\right)
=\mathcal R(\bm G\bm G^\top)
=\mathcal R(\bm G)
=\mathcal R(\bm\eta_{(\ell)}),
\end{aligned}
\]
where
$\mathcal R(\bm G\bm G^\top)=\mathcal R(\bm G)$ follows from
$\mathcal N(\bm G\bm G^\top)=\mathcal N(\bm G^\top)$, and the last
equality follows because
$\bm\Sigma_{(\ell)}^{-1/2}$ is nonsingular.  Using
$\mathcal R(\bm A\otimes\bm B)
=\mathcal R(\bm A)\otimes\mathcal R(\bm B)$, which follows by
writing the columns of $\bm A\otimes\bm B$ as pairwise Kronecker
products of the columns of $\bm A$ and $\bm B$, we obtain
\begin{align}
\mathcal R
\left(
\bm\Phi_0^\top\otimes\bm\eta_{(\ell)}
\right)
=
\mathcal R
\left(
\bm\Delta_0\otimes
\bm\eta_{(\ell)}
\bm\Sigma_{(\ell)}^{-1}
\bm\eta_{(\ell)}^\top
\right).
\label{app:eq:range-equality}
\end{align}

The matrix on the right of \eqref{app:eq:range-equality} is the first
positive semidefinite term in \eqref{app:ch3:T-ell}.  The sum of the
remaining terms is also positive semidefinite because the left-hand
side of \eqref{app:ch3:Phi-information-X-product} is positive
semidefinite, $f_\ell>0$, and $\bm K_{p-q,q}$ is nonsingular.  For
positive semidefinite matrices $\bm P$ and $\bm Q$,
\[
\mathcal N(\bm P+\bm Q)
=
\mathcal N(\bm P)\cap\mathcal N(\bm Q),
\]
and hence
$\mathcal R(\bm P)\subseteq\mathcal R(\bm P+\bm Q)$.
It follows that
\begin{align}
\mathcal R
\left(
\bm\Phi_0^\top\otimes\bm\eta_{(\ell)}
\right)
\nonumber
=
\mathcal R
\left(
\bm\Delta_0\otimes
\bm\eta_{(\ell)}
\bm\Sigma_{(\ell)}^{-1}
\bm\eta_{(\ell)}^\top
\right)
\subseteq
\mathcal R(\bm T_{(\ell)}).
\label{app:eq:range-inclusion}
\end{align}
Finally, $\bm M_{(\ell)}=f_\ell\bm T_{(\ell)}$ and $f_\ell>0$, so
\[
\mathcal R(\bm T_{(\ell)})
=
\mathcal R(\bm M_{(\ell)}).
\]

For an arbitrary vector $\bm a$, set
\[
\bm y
=
(\bm\Phi_0^\top\otimes\bm\eta_{(\ell)})\bm a.
\]
Applying the preceding inequality with
$\bm A=\bm M_{(\ell)}$ and $\bm B=\bm M$ gives
\begin{align}
(\bm\Phi_0\otimes\bm\eta_{(\ell)}^{\top})
\bm M^{\dagger}
(\bm\Phi_0^{\top}\otimes\bm\eta_{(\ell)})
\nonumber
\preceq
(\bm\Phi_0\otimes\bm\eta_{(\ell)}^{\top})
\bm M_{(\ell)}^{\dagger}
(\bm\Phi_0^{\top}\otimes\bm\eta_{(\ell)}).
\label{app:eq:subspace-variance-order}
\end{align}
The first terms on the right-hand sides of \eqref{eq:avar-genv} and
\eqref{eq:avar-sep} are identical.  Adding this common term to both
sides of the displayed inequality gives
\[
\bm V_{(\ell),\mathrm{sep}}
\succeq
\bm V_{(\ell),\mathrm{genv}}.
\]
This proves Corollary~\ref{cor:asymptotic-efficiency}.
\hfill$\square$

\section{Proof of Theorem~\ref{thm:prediction-error-comparison}}
\label{app:proof-prediction}

Fix a group $\ell$, and let
$(\bm X_{(\ell),\mathrm{new}},\bm Y_{(\ell),\mathrm{new}})$ be a
future observation independent of the training sample, where
\begin{equation}
\label{app:eq:new-response}
\bm Y_{(\ell),\mathrm{new}}
=\bm\mu_{(\ell)}
+\bm\beta_{(\ell)}\left(\bm X_{(\ell),\mathrm{new}}-\bm\mu_{\bm X_{(\ell)}}\right)
+\bm\varepsilon_{(\ell),\mathrm{new}}.
\end{equation}
Under joint normality, $\bm\varepsilon_{(\ell),\mathrm{new}}$ is
independent of $\bm X_{(\ell),\mathrm{new}}$, has mean zero, and has
covariance $\bm\Sigma_{(\ell)}$.  Condition on
$\bm X_{(\ell),\mathrm{new}}=\bm x_{\mathrm{new}}$ and write
\begin{equation}
\label{app:eq:x-tilde}
\widetilde{\bm x}
=
\bm x_{\mathrm{new}}
-
\bm\mu_{\bm X_{(\ell)}}.
\end{equation}
By \eqref{app:eq:new-response}, the true conditional mean is
\begin{align}
\bm m_{(\ell)}(\bm x_{\mathrm{new}})
&:=
\operatorname{E}
\left(
\bm Y_{(\ell),\mathrm{new}}
\mid
\bm X_{(\ell),\mathrm{new}}=\bm x_{\mathrm{new}}
\right)
\nonumber\\
&=
\bm\mu_{(\ell)}
+
\bm\beta_{(\ell)}
\left(
\bm x_{\mathrm{new}}-
\bm\mu_{\bm X_{(\ell)}}
\right)
\nonumber\\
&=
\bm\mu_{(\ell)}
+
\bm\beta_{(\ell)}\widetilde{\bm x}.
\label{app:eq:true-conditional-mean}
\end{align}
Consequently,
\begin{equation}
\bm Y_{(\ell),\mathrm{new}}
=\bm m_{(\ell)}(\bm x_{\mathrm{new}})
+\bm\varepsilon_{(\ell),\mathrm{new}}
\qquad
\text{when }\bm X_{(\ell),\mathrm{new}}=\bm x_{\mathrm{new}}.
\end{equation}
For $m\in\{\mathrm{genv},\mathrm{sep}\}$, let
$\widehat{\operatorname{E}}_m$ denote expectation under the model
whose parameters are estimated by method $m$.  Since both methods
estimate the two means by their groupwise sample means,
\begin{align}
\widehat{\bm m}_{(\ell),m}(\bm x_{\mathrm{new}})
&:=
\widehat{\operatorname{E}}_m
\left(
\bm Y_{(\ell),\mathrm{new}}
\mid
\bm X_{(\ell),\mathrm{new}}=\bm x_{\mathrm{new}}
\right)
\nonumber\\
&=
\overline{\bm Y}_{(\ell)}
+
\widehat{\bm\beta}_{(\ell),m}
\left(
\bm x_{\mathrm{new}}-
\overline{\bm X}_{(\ell)}
\right).
\label{app:eq:fitted-conditional-mean}
\end{align}
Define the corresponding conditional prediction-risk matrix by
\begin{align}
\bm R_{(\ell),m}(\bm x_{\mathrm{new}})
&:=\operatorname{E}\left[
\left\{\widehat{\bm m}_{(\ell),m}(\bm x_{\mathrm{new}})-\bm Y_{(\ell),\mathrm{new}}\right\}
\left\{\widehat{\bm m}_{(\ell),m}(\bm x_{\mathrm{new}})-\bm Y_{(\ell),\mathrm{new}}\right\}^{\top}
\,\middle|\,
\bm X_{(\ell),\mathrm{new}}=\bm x_{\mathrm{new}}
\right].
\label{app:eq:prediction-risk-definition}
\end{align}
Here, the expectation is taken over the training sample and the
future response.  Since
\begin{equation}
\widehat{\bm m}_{(\ell),m}(\bm x_{\mathrm{new}})-\bm Y_{(\ell),\mathrm{new}}
=\widehat{\bm m}_{(\ell),m}(\bm x_{\mathrm{new}})-\bm m_{(\ell)}(\bm x_{\mathrm{new}})-\bm\varepsilon_{(\ell),\mathrm{new}},
\end{equation}
the independence and zero mean of the future error give
\begin{align}
\bm R_{(\ell),m}(\bm x_{\mathrm{new}})
&=\bm\Sigma_{(\ell)}
\nonumber\\
&\quad+\operatorname{E}\left[
\left\{\widehat{\bm m}_{(\ell),m}(\bm x_{\mathrm{new}})-\bm m_{(\ell)}(\bm x_{\mathrm{new}})\right\}
\left\{\widehat{\bm m}_{(\ell),m}(\bm x_{\mathrm{new}})-\bm m_{(\ell)}(\bm x_{\mathrm{new}})\right\}^{\top}
\right].
\label{app:eq:prediction-risk-decomposition}
\end{align}
Thus, it remains to calculate the second moment of the
conditional-mean estimation error up to order $n^{-1}$.  Let
\begin{equation}
\label{app:eq:epsilon-bar}
\overline{\bm\varepsilon}_{(\ell)}
=
\frac{1}{n_\ell}
\sum_{j=1}^{n_\ell}\bm\varepsilon_{(\ell)j}.
\end{equation}
Then
\begin{equation}
\overline{\bm Y}_{(\ell)}
=
\bm\mu_{(\ell)}
+
\bm\beta_{(\ell)}
\left(
\overline{\bm X}_{(\ell)}-
\bm\mu_{\bm X_{(\ell)}}
\right)
+
\overline{\bm\varepsilon}_{(\ell)},
\end{equation}
and hence
\begin{equation}
\label{app:eq:conditional-mean-decomposition}
\widehat{\bm m}_{(\ell),m}(\bm x_{\mathrm{new}})-\bm m_{(\ell)}(\bm x_{\mathrm{new}})=\overline{\bm\varepsilon}_{(\ell)}+\left(\widehat{\bm\beta}_{(\ell),m}-\bm\beta_{(\ell)}\right)\widetilde{\bm x}-\left(\widehat{\bm\beta}_{(\ell),m}-\bm\beta_{(\ell)}\right)\left(\overline{\bm X}_{(\ell)}-\bm\mu_{\bm X_{(\ell)}}\right).
\end{equation}
Denote the first two terms on the right-hand side of
\eqref{app:eq:conditional-mean-decomposition} by
$\bm a_{n,m}$ and the last term by $\bm r_{n,m}$, so that
\begin{align}
\bm a_{n,m}
&=
\overline{\bm\varepsilon}_{(\ell)}
+
\left(
\widehat{\bm\beta}_{(\ell),m}
-
\bm\beta_{(\ell)}
\right)
\widetilde{\bm x},
\label{app:eq:a-nm}
\\
\bm r_{n,m}
&=
-
\left(
\widehat{\bm\beta}_{(\ell),m}
-
\bm\beta_{(\ell)}
\right)
\left(
\overline{\bm X}_{(\ell)}
-
\bm\mu_{\bm X_{(\ell)}}
\right).
\label{app:eq:r-nm}
\end{align}
We first show that the terms involving $\bm r_{n,m}$ are negligible
at order $n^{-1}$.  By
Theorems~\ref{thm:genv-asymptotic-distribution} and
\ref{thm:sep-asymptotic-distribution},
\begin{equation}
\label{app:eq:orders-beta-xbar}
\widehat{\bm\beta}_{(\ell),m}
-
\bm\beta_{(\ell)}
=
O_p(n^{-1/2}),
\qquad
\overline{\bm X}_{(\ell)}
-
\bm\mu_{\bm X_{(\ell)}}
=
O_p(n^{-1/2}).
\end{equation}
Hence $\bm r_{n,m}=O_p(n^{-1})$.  By asymptotic normality,
\[
\sqrt n\,\operatorname{vec}\left(\widehat{\bm\beta}_{(\ell),m}-\bm\beta_{(\ell)}\right)
\xrightarrow{d}
\bm Z_{(\ell),m},
\qquad
\bm Z_{(\ell),m}\sim N\left(\bm0,\bm V_{(\ell),m}\right).
\]
The uniform integrability assumption in Theorem~\ref{thm:prediction-error-comparison} permits convergence of the corresponding second moments.  Since $\|\bm A\|_{\mathrm F}=\|\operatorname{vec}(\bm A)\|$, it follows that
\begin{align}
n\operatorname{E}\left\|\widehat{\bm\beta}_{(\ell),m}-\bm\beta_{(\ell)}\right\|_{\mathrm F}^{2}
&=n\operatorname{E}\left\|\operatorname{vec}\left(\widehat{\bm\beta}_{(\ell),m}-\bm\beta_{(\ell)}\right)\right\|^{2}
\nonumber\\
&=\operatorname{E}\left\|\sqrt n\,\operatorname{vec}\left(\widehat{\bm\beta}_{(\ell),m}-\bm\beta_{(\ell)}\right)\right\|^{2}
\nonumber\\
&\longrightarrow \operatorname{E}\left\|\bm Z_{(\ell),m}\right\|^{2}
=\operatorname{E}\operatorname{tr}\left(\bm Z_{(\ell),m}\bm Z_{(\ell),m}^{\top}\right)
\nonumber\\
&=\operatorname{tr}\left\{\operatorname{E}\left(\bm Z_{(\ell),m}\bm Z_{(\ell),m}^{\top}\right)\right\}
=\operatorname{tr}\left(\bm V_{(\ell),m}\right).
\label{app:eq:beta-second-moment}
\end{align}
Also,
\begin{equation}
\operatorname{E}
\left\|
\overline{\bm X}_{(\ell)}-
\bm\mu_{\bm X_{(\ell)}}
\right\|^{2}
=
\frac{1}{n_\ell}
\operatorname{tr}\left(\bm\Sigma_{\bm X_{(\ell)}}\right)
=
O(n^{-1}).
\label{app:eq:xbar-second-moment}
\end{equation}
The independence of the sample means and the centered sample
covariance matrices therefore gives
\begin{align}
\operatorname{E}\|\bm r_{n,m}\|^2
&\leq
\operatorname{E}
\left\|
\widehat{\bm\beta}_{(\ell),m}-\bm\beta_{(\ell)}
\right\|_{\mathrm F}^{2}
\operatorname{E}
\left\|
\overline{\bm X}_{(\ell)}-
\bm\mu_{\bm X_{(\ell)}}
\right\|^{2}
\nonumber\\
&=
O(n^{-1})O(n^{-1})
=
O(n^{-2})
=
o(n^{-1}).
\label{app:eq:remainder-second-moment}
\end{align}
Under joint normality,
$\overline{\bm\varepsilon}_{(\ell)}$ and
$\widehat{\bm\beta}_{(\ell),m}$ are independent.  Since
$\operatorname{E}(\overline{\bm\varepsilon}_{(\ell)})=\bm0$ and
$\operatorname{E}(\overline{\bm\varepsilon}_{(\ell)}
\overline{\bm\varepsilon}_{(\ell)}^\top)
=n^{-1}f_\ell^{-1}\bm\Sigma_{(\ell)}$, the cross terms in
$\operatorname{E}(\bm a_{n,m}\bm a_{n,m}^\top)$ vanish.  Hence
\begin{align}
\operatorname{E}(\bm a_{n,m}\bm a_{n,m}^{\top})
&=\frac{1}{n}f_\ell^{-1}\bm\Sigma_{(\ell)}
+\operatorname{E}\left[
\left(\widehat{\bm\beta}_{(\ell),m}-\bm\beta_{(\ell)}\right)
\widetilde{\bm x}\widetilde{\bm x}^{\top}
\left(\widehat{\bm\beta}_{(\ell),m}-\bm\beta_{(\ell)}\right)^{\top}
\right].
\label{app:eq:a-second-moment-matrix}
\end{align}
Taking the trace of \eqref{app:eq:a-second-moment-matrix} and using
\eqref{app:eq:beta-second-moment} gives
\begin{align*}
\operatorname{E}\|\bm a_{n,m}\|^2
&=\frac{1}{n}f_\ell^{-1}\operatorname{tr}\left(\bm\Sigma_{(\ell)}\right)
+\operatorname{E}\left\|\left(\widehat{\bm\beta}_{(\ell),m}-\bm\beta_{(\ell)}\right)\widetilde{\bm x}\right\|^2
\\
&\leq \frac{1}{n}f_\ell^{-1}\operatorname{tr}\left(\bm\Sigma_{(\ell)}\right)
+\|\widetilde{\bm x}\|^2\operatorname{E}\left\|\widehat{\bm\beta}_{(\ell),m}-\bm\beta_{(\ell)}\right\|_{\mathrm F}^2
=O(n^{-1}).
\end{align*}
The Cauchy--Schwarz inequality then implies
\begin{equation}
\label{app:eq:remainder-cross-moment}
\left\|
\operatorname{E}
(\bm a_{n,m}\bm r_{n,m}^{\top})
\right\|
\leq
\left\{
\operatorname{E}\|\bm a_{n,m}\|^2
\right\}^{1/2}
\left\{
\operatorname{E}\|\bm r_{n,m}\|^2
\right\}^{1/2}
=
o(n^{-1}).
\end{equation}
The transpose cross moment is also $o(n^{-1})$.

Combining \eqref{app:eq:a-second-moment-matrix},
\eqref{app:eq:remainder-second-moment}, and
\eqref{app:eq:remainder-cross-moment} gives
\begin{align}
&\operatorname{E}
\left[
\left\{
\widehat{\bm m}_{(\ell),m}(\bm x_{\mathrm{new}})
-
\bm m_{(\ell)}(\bm x_{\mathrm{new}})
\right\}
\left\{
\widehat{\bm m}_{(\ell),m}(\bm x_{\mathrm{new}})
-
\bm m_{(\ell)}(\bm x_{\mathrm{new}})
\right\}^{\top}
\right]
\nonumber\\
&\quad=
\frac{1}{n}f_\ell^{-1}\bm\Sigma_{(\ell)}
+
\operatorname{E}
\left[
\left(
\widehat{\bm\beta}_{(\ell),m}-\bm\beta_{(\ell)}
\right)
\widetilde{\bm x}\widetilde{\bm x}^{\top}
\left(
\widehat{\bm\beta}_{(\ell),m}-\bm\beta_{(\ell)}
\right)^{\top}
\right]
+
o(n^{-1}).
\label{app:eq:mean-error-second-moment-decomposition}
\end{align}
Thus, it remains to evaluate the second term on the right-hand side.
The asymptotic covariance $\bm V_{(\ell),m}$ is given for
$\operatorname{vec}(\widehat{\bm\beta}_{(\ell),m})$.  To apply it to
that term, define
\begin{equation}
\label{app:eq:H-x}
\bm H(\widetilde{\bm x})
=
\widetilde{\bm x}^{\top}\otimes\bm I_r.
\end{equation}
Then
\begin{equation}
\label{app:eq:vec-prediction-identity}
\left(
\widehat{\bm\beta}_{(\ell),m}
-
\bm\beta_{(\ell)}
\right)
\widetilde{\bm x}
=
\bm H(\widetilde{\bm x})
\operatorname{vec}
\left(
\widehat{\bm\beta}_{(\ell),m}
-
\bm\beta_{(\ell)}
\right).
\end{equation}
Asymptotic normality and uniform integrability then give
\begin{equation}
\operatorname{E}\left[\left(\widehat{\bm\beta}_{(\ell),m}-\bm\beta_{(\ell)}\right)\widetilde{\bm x}\widetilde{\bm x}^{\top}\left(\widehat{\bm\beta}_{(\ell),m}-\bm\beta_{(\ell)}\right)^{\top}\right]
=\frac{1}{n}\bm H(\widetilde{\bm x})\bm V_{(\ell),m}\bm H(\widetilde{\bm x})^{\top}+o(n^{-1}).
\label{app:eq:coefficient-prediction-moment}
\end{equation}
Here, $\bm V_{(\ell),m}$ is given by \eqref{eq:avar-genv} or
\eqref{eq:avar-sep}.

Substituting \eqref{eq:avar-genv} and using the mixed-product rule for
Kronecker products gives
\begin{align}
\bm H(\widetilde{\bm x})\bm V_{(\ell),\mathrm{genv}}\bm H(\widetilde{\bm x})^{\top}
&=f_\ell^{-1}\bm H(\widetilde{\bm x})\left(\bm\Phi\bm\Delta_{(\ell)}^{-1}\bm\Phi^{\top}\otimes\bm\Sigma_{(\ell)}\right)\bm H(\widetilde{\bm x})^{\top}
\nonumber\\
&\quad+\bm H(\widetilde{\bm x})(\bm\Phi_0\otimes\bm\eta_{(\ell)}^{\top})\bm M^\dagger(\bm\Phi_0^{\top}\otimes\bm\eta_{(\ell)})\bm H(\widetilde{\bm x})^{\top}
\nonumber\\
&=f_\ell^{-1}\left(\widetilde{\bm x}^{\top}\bm\Phi\bm\Delta_{(\ell)}^{-1}\bm\Phi^{\top}\widetilde{\bm x}\right)\bm\Sigma_{(\ell)}
\nonumber\\
&\quad+\left(\widetilde{\bm x}^{\top}\bm\Phi_0\otimes\bm\eta_{(\ell)}^{\top}\right)\bm M^\dagger\left(\bm\Phi_0^{\top}\widetilde{\bm x}\otimes\bm\eta_{(\ell)}\right).
\label{app:eq:genv-coefficient-prediction}
\end{align}
Combining \eqref{app:eq:mean-error-second-moment-decomposition},
\eqref{app:eq:coefficient-prediction-moment}, and
\eqref{app:eq:genv-coefficient-prediction} gives
\begin{align}
&
\operatorname{E}
\left[
\left\{
\widehat{\bm m}_{(\ell),\mathrm{genv}}(\bm x_{\mathrm{new}})
-
\bm m_{(\ell)}(\bm x_{\mathrm{new}})
\right\}
\left\{
\widehat{\bm m}_{(\ell),\mathrm{genv}}(\bm x_{\mathrm{new}})
-
\bm m_{(\ell)}(\bm x_{\mathrm{new}})
\right\}^{\top}
\right]
\nonumber\\
&=
\frac{1}{n}f_\ell^{-1}
\left\{
1
+
\widetilde{\bm x}^{\top}
\bm\Phi\bm\Delta_{(\ell)}^{-1}\bm\Phi^{\top}
\widetilde{\bm x}
\right\}
\bm\Sigma_{(\ell)}
\nonumber\\
&\quad+
\frac{1}{n}
\left(
\widetilde{\bm x}^{\top}\bm\Phi_0
\otimes
\bm\eta_{(\ell)}^{\top}
\right)
\bm M^\dagger
\left(
\bm\Phi_0^{\top}\widetilde{\bm x}
\otimes
\bm\eta_{(\ell)}
\right)
+
o(n^{-1}).
\label{app:eq:mean-risk-genv}
\end{align}
Applying \eqref{app:eq:prediction-risk-decomposition} to
\eqref{app:eq:mean-risk-genv} gives
\begin{align}
\bm R_{(\ell),\mathrm{genv}}(\bm x_{\mathrm{new}})
={}&\bm\Sigma_{(\ell)}
+\frac{1}{n}f_\ell^{-1}
\left\{1+\widetilde{\bm x}^{\top}\bm\Phi\bm\Delta_{(\ell)}^{-1}\bm\Phi^{\top}\widetilde{\bm x}\right\}
\bm\Sigma_{(\ell)}
\nonumber\\
&+\frac{1}{n}
\left(\widetilde{\bm x}^{\top}\bm\Phi_0\otimes\bm\eta_{(\ell)}^{\top}\right)
\bm M^\dagger
\left(\bm\Phi_0^{\top}\widetilde{\bm x}\otimes\bm\eta_{(\ell)}\right)
+o(n^{-1}).
\label{app:eq:risk-genv-derived}
\end{align}

The same calculation using \eqref{eq:avar-sep} yields
\begin{align}
\bm R_{(\ell),\mathrm{sep}}(\bm x_{\mathrm{new}})
={}&\bm\Sigma_{(\ell)}
+\frac{1}{n}f_\ell^{-1}
\left\{1+\widetilde{\bm x}^{\top}\bm\Phi\bm\Delta_{(\ell)}^{-1}\bm\Phi^{\top}\widetilde{\bm x}\right\}
\bm\Sigma_{(\ell)}
\nonumber\\
&+\frac{1}{n}
\left(\widetilde{\bm x}^{\top}\bm\Phi_0\otimes\bm\eta_{(\ell)}^{\top}\right)
\bm M_{(\ell)}^\dagger
\left(\bm\Phi_0^{\top}\widetilde{\bm x}\otimes\bm\eta_{(\ell)}\right)
+o(n^{-1}).
\label{app:eq:risk-separate-derived}
\end{align}
Subtracting \eqref{app:eq:risk-genv-derived} from
\eqref{app:eq:risk-separate-derived} gives
\begin{align}
&
n
\left\{
\bm R_{(\ell),\mathrm{sep}}(\bm x_{\mathrm{new}})
-
\bm R_{(\ell),\mathrm{genv}}(\bm x_{\mathrm{new}})
\right\}
\nonumber\\
&\quad=
\left(
\widetilde{\bm x}^{\top}\bm\Phi_0
\otimes
\bm\eta_{(\ell)}^{\top}
\right)
\left(
\bm M_{(\ell)}^\dagger-\bm M^\dagger
\right)
\left(
\bm\Phi_0^{\top}\widetilde{\bm x}
\otimes
\bm\eta_{(\ell)}
\right)
+
o(1).
\label{app:eq:risk-difference}
\end{align}
Using $\bm H(\widetilde{\bm x})=\widetilde{\bm x}^{\top}\otimes\bm I_r$,
\eqref{eq:avar-genv}, \eqref{eq:avar-sep}, and the mixed-product rule,
the leading matrix in \eqref{app:eq:risk-difference} can be written as
\begin{align*}
&\left(\widetilde{\bm x}^{\top}\bm\Phi_0\otimes\bm\eta_{(\ell)}^{\top}\right)
\left(\bm M_{(\ell)}^\dagger-\bm M^\dagger\right)
\left(\bm\Phi_0^{\top}\widetilde{\bm x}\otimes\bm\eta_{(\ell)}\right)
\\
&\qquad=\bm H(\widetilde{\bm x})
\left(\bm V_{(\ell),\mathrm{sep}}-\bm V_{(\ell),\mathrm{genv}}\right)
\bm H(\widetilde{\bm x})^{\top}.
\end{align*}
By Corollary~\ref{cor:asymptotic-efficiency},
$\bm V_{(\ell),\mathrm{sep}}-\bm V_{(\ell),\mathrm{genv}}\succeq\bm0$,
so the right-hand side, and hence the leading matrix, is positive
semidefinite.  It is precisely
$\bm B_{(\ell)}(\bm x_{\mathrm{new}})$ defined in the theorem, so
\[
n
\left\{
\bm R_{(\ell),\mathrm{sep}}(\bm x_{\mathrm{new}})
-
\bm R_{(\ell),\mathrm{genv}}(\bm x_{\mathrm{new}})
\right\}
\longrightarrow
\bm B_{(\ell)}(\bm x_{\mathrm{new}})
\succeq
\bm 0.
\]
This proves Theorem~\ref{thm:prediction-error-comparison}.
\hfill $\square$

\end{document}